\documentclass[11pt,a4paper]{article}

\pdfoutput=1
\usepackage{jheppub}
\usepackage{graphicx,color,verbatim,epsfig,subfigure}
\usepackage{latexsym,amsfonts,amssymb,amsmath,makeidx,mathrsfs,multirow,bm,slashed,fancyvrb}
\usepackage[makeroom]{cancel}
\usepackage[utf8]{inputenc}
\usepackage{comment}

\def\beqn{\begin{eqnarray}}
\def\eeqn{\end{eqnarray}}
\def\beqs{\begin{subequations}}
\def\eeqs{\end{subequations}}
\def\beq{\begin{equation}}
\def\eeq{\end{equation}}
\def\ba{\begin{array}}
\def\ea{\end{array}}

\def\non{\nonumber\\}

\def\hf{\frac{1}{2}}
\def\[{\left[}
\def\]{\right]}
\def\({\left(}
\def\){\right)}

\def\TeV{\rm TeV}
\def\GeV{\rm GeV}
\def\MeV{\rm MeV}

\def\gU{\rm U}
\def\gSU{\rm SU}

\def\mA{\mathcal{A}}

\def\mG{\mathcal{G}}
\def\mH{\mathcal{H}}

\def\mL{\mathcal{L}}
\def\mM{\mathcal{M}}

\def\mO{\mathcal{O}}

\def\mR{\mathcal{R}}

\preprint{}

\title{Collapsing domain walls in the two-Higgs-doublet model and deep insights from the EDM}

\author[\alpha]{~Ning Chen }
\emailAdd{chenning$\_$symmetry@nankai.edu.cn}
\author[\alpha]{~Tong Li }
\emailAdd{litong@nankai.edu.cn}
\author[\alpha]{~Zhaolong Teng}
\emailAdd{tengcl@mail.nankai.edu.cn}
\affiliation[\alpha]{School of Physics, Nankai University, Tianjin 300071, China}
\author[\beta]{~Yongcheng Wu}
\emailAdd{ycwu@physics.carleton.ca}
\affiliation[\beta]{Ottawa-Carleton Institute for Physics, \\
Carleton University, 1125 Colonel By Drive, Ottawa, Ontario K1S 5B6, Canada }

\abstract
{\\[1mm]
We study the domain wall solutions in the general two-Higgs-doublet model (2HDM) with a CP-violating phase.
The 2HDM with the spontaneouse CP violation is found to have domain wall solutions whose tensions are $\mathcal{O}(10^6)\,{\rm GeV}^3$, which are excluded by the Zel'dovich-Kobzarev-Okun bound.
With the explicit CP-violating (CPV) terms as the so-called biased term in the scalar potential, domain walls can collapse in the early Universe.
The sizes of the explicit CP violation can be constrained from the Big Bang nucleosynthesis.
This constraint is converted to the CPV mixing of $\alpha_c$, and is mostly sensitive to the mass splittings between two heavy neutral Higgs bosons.
We estimate the possible gravitational wave signals and the electric dipole moment (EDM) predictions due to the domain wall collapsing.
It turns out that the peak spectrum of the GW from the domain wall collapsing cannot be probed in any future program.
In contrast, the untenable regions with very tiny explicit CPV parameter in the Higgs potential has been partially excluded by the latest electron EDM measurements at the ACME-II and will be further confirmed or excluded by the future ACME-III projection.
}

\arxivnumber{arXiv:2006.06913}

\begin{document}

\maketitle
\setcounter{page}{2}

\newpage


\section{Introduction}
\label{section:intro}

The Standard Model (SM) of particle physics has been experimentally verified to be successful, with the discovery of the $125\,\GeV$ Higgs boson~\cite{Aad:2012tfa,Chatrchyan:2012ufa} in 2012.
Yet, the SM itself cannot address the long-lasting puzzle of the baryon asymmetry in the Universe (BAU).
Thus, many efforts have been made to realize the three Sakharov conditions~\cite{Sakharov:1967dj} by extending the SM.
It is quite general that the extensions to the SM lead to larger symmetries.
During the phase transitions of the early Universe, different symmetry breaking patterns are likely to produce topological defects.
This is known as the Kibble-Zurek mechanism~\cite{Kibble:1976sj,Zurek:1985qw,Vachaspati:2006zz}.
The domain walls, which are two-dimensional topological defects, can exist when the $0$-th homotopy group associated with the symmetry breaking of $\mG \to \mH$ is nontrivial, i.e., $\Pi_0(\mG/\mH) \neq \mathbf{1}$.
The existence of the domain wall solutions in any new physics can be problematic, since their energy density scale as $t^{-1}$ with respect to time $t$.
This means once they are formed in the early Universe, they can dominate the energy densities over radiation and matter and thus spoil the standard cosmology.

In this work, we study the possible domain wall solutions arising from the CP symmetries of the two-Higgs-doublet model (2HDM).
Originally, the study of the electroweak theory with two Higgs doublets in the scalar sector was motivated to achieve the spontaneous CP violations (SCPV)~\cite{Lee:1974jb}.
However, the existence of two degenerate vacua with opposite-sign CP phases can lead to the domain wall solutions.
The domain wall solutions in the 2HDM were previously studied in Refs.~\cite{Bao:2009sa,Ipek:2013iba,Grzadkowski:2014ada}, and recently revisited in Refs.~\cite{Eto:2018hhg,Eto:2018tnk}.
With the typical energy scale of few hundred GeV for the heavy scalars, the tensions of the stable domain walls in the 2HDM are found of the size $\sigma \sim \mO(10^6)\,\GeV^3$, which is excluded by the Zel'dovich-Kobzarev-Okun bound~\cite{Zeldovich:1974uw}.

Given the possible domain wall solutions in the CPV 2HDM, let us briefly review the existing solutions.
The first well-known method in the new physics model building is to invoke an inflation phase~\cite{Guth:1980zm} during the domain wall formation.
However, the inflation models typically operate at the GUT scale of $\Lambda_{\rm GUT}\sim 10^{16}\,\GeV$, which does not seem to remedy the SCPV domain wall forming at the electroweak scale. Note that there exist other possibilities to solve the domain wall problem if some specific global symmetry is spontaneously broken in a post inflation epoch. 
For the Peccei-Quinn symmetry solving the strong CP problem in QCD, one solution is the so-called Lazarides-Shafi mechanism~\cite{Lazarides:1982tw} in which the degenerate vacua are connected to each other by another continuous group transformation. 
The choice of the symmetry group and Higgs representations in this mechanism has to be cautious, which makes the possibility difficult to happen~\cite{Barr:1982bb}. 
If the global lepton number symmetry is spontaneously broken, one can introduce auxiliary Majorana field with non-trivial $\gSU(2)_L$ quantum number in Majoron model or a supersymmetric context to solve the domain wall problem~\cite{Lazarides:2018aev}.
One possible way to evade the domain wall problem is to include a small symmetry breaking term, that is the so-called biased term in the Higgs potential.
In this perspective, the symmetry of the CP transformation is approximate.
With such a mechanism~\cite{Vilenkin:1981zs,Gelmini:1988sf,Larsson:1996sp}, the possible domain walls were unstable and collapsed before they overclose the Universe.
Indeed, we found that such biased terms are nothing but the explicit CPV (ECPV) parameters in the 2HDM potential.

With the ECPV parameters as the biased terms in the 2HDM potential, one may expect gravitational wave (GW) signals associated with the domain wall collapses.
This was previously discussed in Refs.~\cite{Hiramatsu:2010yz,Kawasaki:2011vv,Saikawa:2017hiv,Zhou:2020ojf,Chen:2020wvu,Jaeckel:2020mqa}.
Therefore, it is natural to ask if such GW signals can be probed at the future programs, such as the satellite-based interferometers of LISA~\cite{AmaroSeoane:2012km,AmaroSeoane:2012je}, Taiji~\cite{Guo:2018npi}, and Tianqin~\cite{Luo:2015ght}, or the radio telescope of square kilometer arrays (SKA)~\cite{Janssen:2014dka} and the Japanese space GW antenna (DECIGO)~\cite{Kawamura:2011zz}.
Note that such GW signals are different from those produced due to the bubble collisions during the strongly first-order phase transitions.
The constraint on the ECPV parameters is imposed so that the corresponding biased terms are sufficiently large to collapse the domain walls before the epoch of the the big bang nucleosynthesis (BBN).
In Ref.~\cite{Chen:2020wvu}, we found such signals can be sufficiently strong when the corresponding new physics scale is above about $10\,\TeV$.
However, as we shall show, the related GW peak spectrum here is typically $\lesssim \mO(10^{-23})$ and below the future search sensitivities of SKA~\cite{Janssen:2014dka} and DECIGO~\cite{Kawamura:2011zz}.
Alternatively, the precise measurements of the electric dipole moments (EDM) play an intriguing role in constraining the small ECPV parameters.
The ongoing and upcoming EDM measurements include the electron EDM (eEDM) from the ACME collaboration~\cite{Baron:2013eja,Andreev:2018ayy}, the mercury EDM~\cite{Graner:2016ses}, and the radium EDM~\cite{Bishof:2016uqx}.
The EDM measurements are usually interpreted to constrain CPV mixings in the context of the 2HDM~\cite{Shu:2013uua,Inoue:2014nva,Bian:2014zka,Chen:2015gaa,Bian:2016zba,Egana-Ugrinovic:2018fpy,Chun:2019oix,Cheung:2020ugr,Kanemura:2020ibp}, or the SUSY-breaking scale~\cite{Cesarotti:2018huy}. Usually, the constraint on CP-violating 2HDM from EDM was performed in the absence of spontaneous CP-violating phase to avoid the problematic SCPV domain walls~\cite{Inoue:2014nva}.
With the future improvements of the eEDM precision measurements from the ACME-III, given a few hundred GeV heavy scalar masses, we find that they can be used to exclude sizable regions of the ECPV parameters in the 2HDM potential jointly with the BBN constraint.
\textit{In this sense, the EDM experiments provide us deep insights to the very tiny ECPV term that triggered the domain wall collapse in the early Universe.}
This is completely different from what was usually discussed in the context of SUSY, such that the precision measurements of the EDM will constrain the corresponding new physics scale above $\mO(10)\,\TeV$~\cite{Cesarotti:2018huy}.

The rest of the paper is organized as follows.
In Sec.~\ref{section:2HDM}, we review the general 2HDM.
We focus on the minimization conditions and the mass spectrum in both the SCPV and the ECPV scenarios.
The parameters between the generic basis and the physical basis are built for solving the domain wall.
The theoretical constraints of the perturbative unitarity and the vacuum stabilities are imposed to the scalar self couplings.
In Sec.~\ref{section:2HDMDW}, we present the domain wall solutions in the CPV 2HDM.
The domain walls from the SCPV typically have tensions of $\mO(10^6)\,\GeV^3$, which should be collapsed with sufficiently sizable biased terms.
The ECPV parameter in the 2HDM, which we choose to be ${\rm Im}\lambda_5$, will be bounded from below by considering the BBN constraint.
By estimating the corresponding GW signals from the domain wall collapses, they are beyond the search sensitivities of any future satellite-based observation programs.
In Sec.~\ref{section:EDM}, we turn to the estimations of the eEDM in the domain wall collapsing scenario.
The current and the future projections of the eEDM measurements from the ACME are used to set upper limits to the CPV mixing angle, as well as exclude the ECPV parameter directly.
We summarize our results in Sec.~\ref{section:conclusion}.
In the appendix~\ref{section:2HDMspectrum}, we provide details of deriving the 2HDM mass spectrum for both the SCPV and the ECPV scenarios. 
Besides, the Yukawa couplings for the CPV 2HDM are also presented therein.


\section{The general 2HDM with the CPV}
\label{section:2HDM}

In this section, we review the general CP-violating 2HDM and focus on the global symmetry from the CP transformation of the potential.
Previous studies of the symmetries and topological structures of the 2HDM include Refs.~\cite{Davidson:2005cw,Gunion:2005ja,Ivanov:2006yq,Ivanov:2007de,Branco:2011iw,Battye:2011jj,Grzadkowski:2013rza,Inoue:2014nva,Eto:2018hhg}.
Our focus will be on the CP symmetries in the 2HDM potential, with the details of deriving the relative CP-violating phase of $\theta$ given in Sec.~\ref{section:2HDMSCPV} and \ref{section:2HDMECPV} for the SCPV and the ECPV scenarios, respectively.
The Yukawa couplings for the CPV 2HDM are listed in Sec.~\ref{section:2HDM_Yukawa}.

\subsection{The 2HDM potential}

We write down the 2HDM potential with the softly broken $\mathbb{Z}_2$ symmetry as
\beqn\label{eq:2HDMpotential}
V(\Phi_1\,, \Phi_2)&=&m_{11}^2 |\Phi_1|^2 + m_{22}^2 |\Phi_2|^2 - ( m_{12}^2 \Phi_1^\dag \Phi_2 + h.c.) \non
&+ & \frac{\lambda_1}{2} |\Phi_1 |^4 + \frac{\lambda_2}{2} | \Phi_2|^4 + \lambda_3 |\Phi_1 |^2  | \Phi_2 |^2 \non
&+& \lambda_4 (\Phi_1^\dag \Phi_2) (\Phi_2^\dag \Phi_1) +  \Big[  \frac{\lambda_5}{2} ( \Phi_1^\dag \Phi_2 )^2  + h.c. \Big]\,.
\eeqn
Here, $(m_{12}^2\,, \lambda_5)$ are complex for the CPV case, and we parametrize them by $m_{12}^2\equiv | m_{12}^2 | e^{ i \delta_1} = {\rm Re} m_{12}^2+ i {\rm Im} m_{12}^2  $ and $\lambda_5\equiv |\lambda_5| e^{i\delta_2} = {\rm Re} \lambda_5 + i {\rm Im} \lambda_5 $.
All parameters are real for the CP-conserving (CPC) case.
To study the domain wall solution of the 2HDM under the CP transformations below, we follow the Refs.~\cite{Grzadkowski:2009bt,Grzadkowski:2013rza,Grzadkowski:2014ada} to parametrize two Higgs doublets as follows
\beqn
&& \Phi_1 =  \left( \ba{c} H_1^+  \\  \frac{1}{ \sqrt{2} }( \varphi_1 + H_1^0 + i A_1^0)    \ea \right) \,,\quad  \Phi_2 = e^{i \Theta }  \left( \ba{c} H_2^+  \\  \frac{1}{ \sqrt{2} }( \varphi_2    + H_2^0 + i A_2^0  ) \ea \right) \,.
\eeqn
In order to obtain the domain wall solutions below, we treat $(\varphi_1\,,\varphi_2\,,\Theta)$ as background fields.
When the electroweak symmetry breaking (EWSB) occurred, they obtain the vacuum expectation values (VEVs) as $\langle \varphi_{1\,,2} \rangle = v_{1\,,2}$, $\langle \Theta \rangle = \theta$, with $\theta$ to be solved below for both the SCPV and the ECPV scenarios.
As usual, two VEVs are parametrized as $ v_1 = v c_\beta$ and $ v_2 = v s_\beta$~\footnote{Throughout the context, we use the short-handed notations of $s_\beta \equiv \sin\beta$ and $c_\beta \equiv \cos\beta$.}.
The minimizations of the most general 2HDM poential~\eqref{eq:2HDMpotential} are the following
\beqs\label{eqs:ECPV_min}
\beqn
&&m_{11}^2= \frac{ t_\beta }{ c_\theta } {\rm Re} m_{12}^2 - \hf \lambda_1 v^2 c_\beta^2 - \hf ( \lambda_3 + \lambda_4 {\color{magenta}+}  {\rm Re} \lambda_5 {\color{magenta}-} {\rm Im} \lambda_5 t_\theta ) v^2 s_\beta^2  \,,\label{eq:ECPV_min01}\\
&& m_{22}^2= \frac{1}{ t_\beta c_\theta } {\rm Re} m_{12}^2  - \hf \lambda_2 v^2 s_\beta^2 - \hf ( \lambda_3 + \lambda_4 + {\rm Re} \lambda_5 - {\rm Im} \lambda_5 t_\theta  ) v^2 c_\beta^2  \,,\label{eq:ECPV_min02}\\
&&{\rm Re } m_{12}^2 s_\theta + {\rm Im} m_{12}^2 c_\theta =   \frac{1}{ 2 } ( {\rm Re} \lambda_5 s_{2\theta} + {\rm Im} \lambda_5 c_{2\theta} ) v^2  s_\beta c_\beta \,.\label{eq:ECPV_min03}
\eeqn
\eeqs
In the special case of vanishing spontaneous CPV phase $\theta =0$, the minimization conditions are reduced to
\beqs\label{eqs:ECPVnoSCPV_min}
\beqn
m_{11}^2&=& {\rm Re} m_{12}^2 t_\beta - \hf \lambda_1 v^2 c_\beta^2 - \hf ( \lambda_3 + \lambda_4 +  {\rm Re} \lambda_5  ) v^2 s_\beta^2  \,,\label{eq:ECPVnoSCPV_min01}\\
m_{22}^2&=& {\rm Re} m_{12}^2 /t_\beta  - \hf \lambda_2 v^2 s_\beta^2 - \hf ( \lambda_3 + \lambda_4 + {\rm Re} \lambda_5  ) v^2 c_\beta^2  \,,\label{eq:ECPVnoSCPV_min02}\\
{\rm Im} m_{12}^2&=&   \frac{1}{ 2 } {\rm Im}\lambda_5  v^2 s_\beta c_\beta \,.\label{eq:ECPVnoSCPV_min03}
\eeqn
\eeqs

\subsection{The 2HDM potential with the SCPV}

In a general 2HDM, the CP symmetry is conserved only if three invariants of $J_{1\,,2\,,3}$ defined in Refs.~\cite{Davidson:2005cw,Gunion:2005ja,Grzadkowski:2013rza} are all real.
This can be realized with two possibilities, namely, either with
\beqn\label{eq:SCPV1}
{\rm SCPV1}&:& {\rm Im}[ ( m_{12}^2 )^2 \lambda_5^*] =0\,,
\eeqn
or
\beqn\label{eq:SCPV2}
{\rm SCPV2}&:& \lambda_1=\lambda_2\,,\qquad m_{11}^2=m_{22}^2\,,\qquad t_\beta =1\,.
\eeqn
In this work, we focus on the SCPV1 scenario defined in Ref.~\cite{Grzadkowski:2013rza}, and use the notion of SCPV throughout the context.
Thus, it is equivalent to take $2 \delta_1 - \delta_2 = n \pi$, with $n$ being an integer.
Without loss of generality, we shall take $n=0$ for our later discussions.
In order to simplify the calculation, we make a rephase of $\Phi_2 \rightarrow e^{- i \delta_1} \Phi_2$ to eliminate the phases in the coefficients, so that ${\rm Im} m_{12}^2  = {\rm Im} \lambda_5 = 0$.
The minimization conditions for the SCPV scenario are then expressed as
\beqs\label{eqs:SCPV_min}
\beqn
 m_{11}^2 &=&   -\hf \lambda_1 v^2 c_\beta^2 - \hf ( \lambda_3 + \lambda_ 4 - {\rm Re} \lambda_5)  v^2 s_\beta^2 \,,\label{eq:SCPV1_min01}\\
 m_{22}^2 &=& - \hf \lambda_2 v^2 s_\beta^2 - \hf ( \lambda_3 + \lambda_ 4 - {\rm Re} \lambda_5 )v^2 c_\beta^2 \,,\label{eq:SCPV1_min02}\\
{\rm Re} m_{12}^2  &=& {\rm Re} \lambda_5  v^2  s_\beta c_\beta  c_{\theta}  \,.\label{eq:SCPV1_min03}
\eeqn
\eeqs
Two degenerate solutions of
\beqn
 \theta&=& \pm \cos^{-1} \Big( \frac{ {\rm Re} m_{12}^2  }{ {\rm Re} \lambda_5 v^2  s_\beta c_\beta }  \Big)
\eeqn
with two opposite signs lead to the domain wall solutions.
Note that this expression of $\theta$ is not our solution in the physical basis.
With the minimization conditions of the SCPV scenario in Eqs.~\eqref{eqs:SCPV_min}, the 2HDM potential can be expressed as
\beqs\label{eqs:SCPV1_potential}
\beqn
V&=& V_{\rm CPC} + V_{\rm SCPV} \,,\\
V_{\rm CPC}&=& \hf \Big( m_{11}^2 c_\beta^2  + m_{22}^2 s_\beta^2 \Big) v^2 + \Big[  \frac{ 1 }{ 8 } \lambda_1 c_\beta^4  + \frac{1 }{ 8 } \lambda_2 s_\beta^4   +\frac{1}{4} ( \lambda_3 + \lambda_4  ) s_\beta^2 c_\beta^2  \Big]  v^4 \,,\\
V_{\rm SCPV}&=& -  {\rm Re} m_{12}^2   c_{\theta } s_\beta c_\beta  v^2 +\frac{1}{4}  {\rm Re} \lambda_5  c_{2\theta }  s_\beta^2 c_\beta^2  v^4  = - \frac{1}{4} {\rm Re} \lambda_5  (1 +2 c_{\theta }^2 )  s_\beta^2  c_\beta^2 v^4\,.
\eeqn
\eeqs
This is consistent with what was obtained in Ref.~\cite{Eto:2018hhg}.
The parameters in the Higgs potential~\eqref{eqs:SCPV1_potential} will be used for estimating the domain wall tension and the bound to the collapses later. 
They will be obtained by using the inputs from the physical basis, and the readers should refer to Sec.~\ref{section:2HDMSCPV} for details.

\begin{figure}[htb]
\centering
\includegraphics[height=5cm]{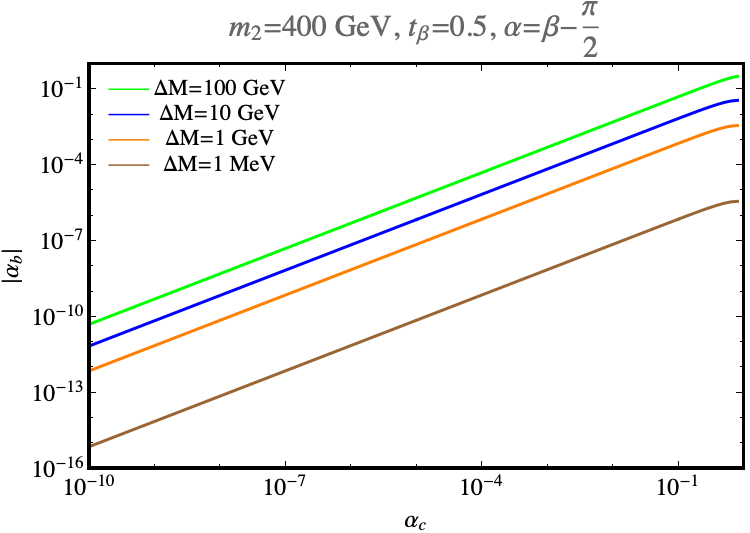}
\includegraphics[height=5cm]{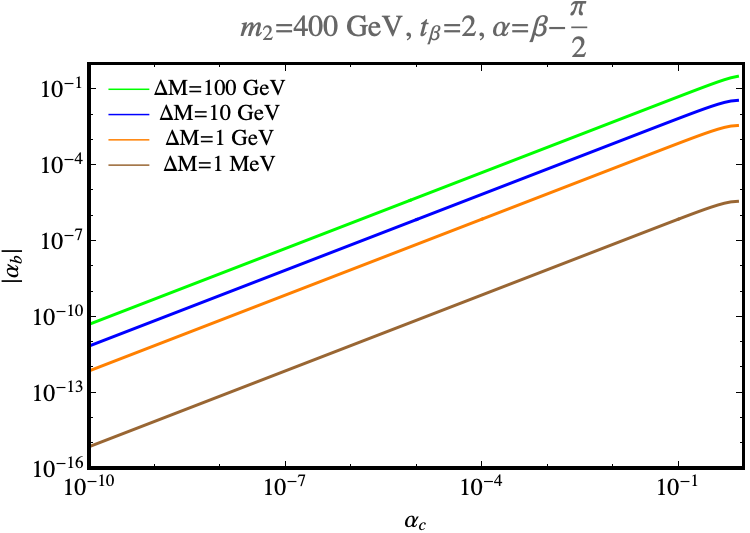}
\caption{\label{fig:alphab}
The values of $|\alpha_b|$ versus the $\alpha_c$ with various inputs of $\Delta M = m_3 - m_2 = (100\,\GeV\,, 10\,\GeV\,, 1\,\GeV\,, 1\,\MeV)$, with $t_\beta=0.5$ (left) and $t_\beta=2$ (right), and $\alpha=\beta-\frac{\pi}{2}$.
}
\end{figure}

In both the SCPV and the ECPV scenarios, the CPV mixing angles of $\alpha_b$ and $\alpha_c$ are related to each other according to Eq.~\eqref{eq:mass_constraint}.
Without loss of generality, we take $\alpha_c$ as the input parameter.
In Fig.~\ref{fig:alphab}, we display the values of $|\alpha_b|$ versus the input CPV mixing angle of $\alpha_c\in (10^{-10}\,, 1)$ with various mass splittings of $\Delta M= m_3 - m_2$ between two heavy neutral Higgs bosons.
The decreasing values of $\alpha_c$ also lead to smaller values of $|\alpha_b |$.
Also, the sizes of the CPV mixing angles are suppressed with very degenerate mass splitting of $\Delta M$, as we vary $\Delta M$ from $100\,\GeV$ to $1\,\MeV$ in the plots.
This pattern can be found straightforwardly from Eq.~\eqref{eq:alphab_Approx} as well.
Therefore, the joint effects of smaller input parameter $\alpha_c$ and the smaller mass splitting of $\Delta M$ can lead to suppressions to the eEDM.
Note that the relations of $ \alpha_b$ versus $\alpha_c$ as given in Eq.~\eqref{eq:mass_constraint} and Fig.~\ref{fig:alphab} are independent of the quartic scalar self coupling terms in the 2HDM potential, thus they hold for both the SCPV and the ECPV scenarios.
Since the $\alpha_b$ dependences on $\alpha_c$ and the mass splitting $\Delta M$ are close between the $t_\beta =0.5$ and $t_\beta=2$ cases, we shall always use the $t_\beta =0.5$ as our benchmark throughout our discussions below.

\subsection{The 2HDM potential with the ECPV}

\begin{figure}[htb]
\centering
\includegraphics[height=5cm]{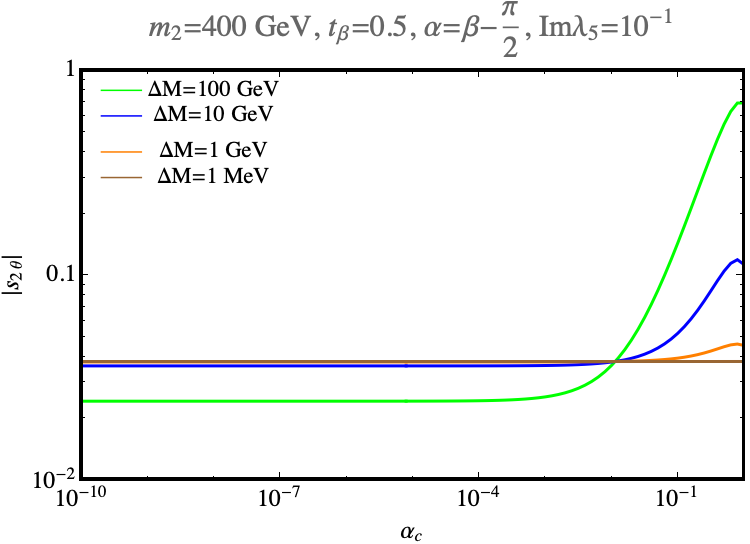}
\includegraphics[height=5cm]{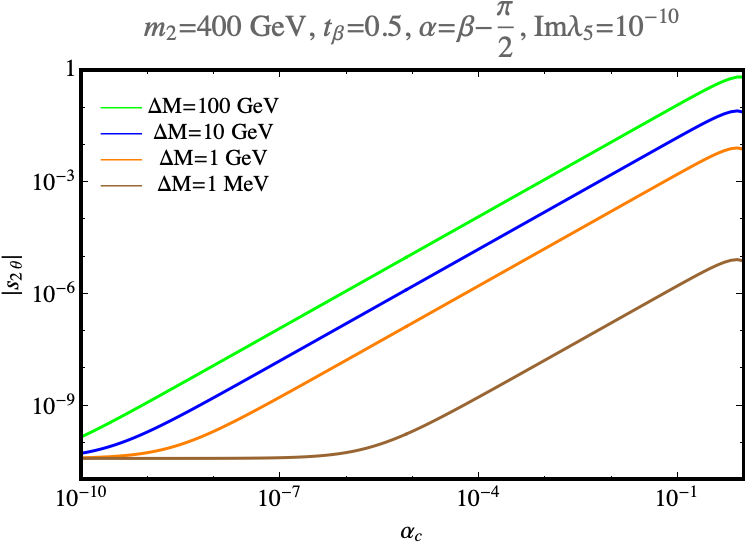}\\
\includegraphics[height=5cm]{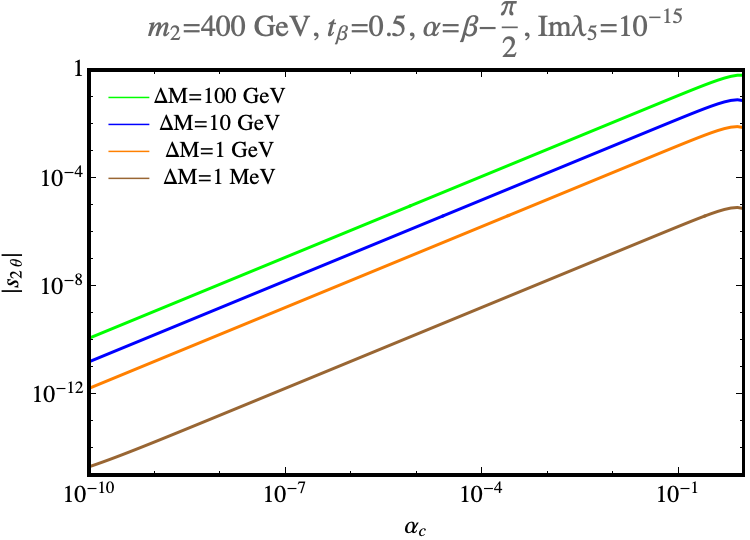}
\includegraphics[height=5cm]{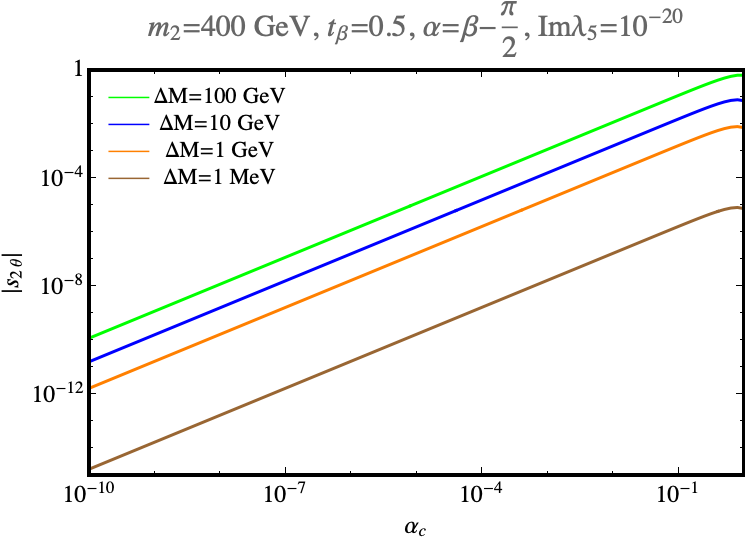}
\caption{\label{fig:ECPV_theta}
The values of $|s_{2\theta}|$ versus the $\alpha_c$ with various inputs of $\Delta M = m_3 - m_2 = (100\,\GeV\,, 10\,\GeV\,, 1\,\GeV\,, 1\,\MeV)$, with $\tan\beta=0.5$ and $\alpha=\beta-\frac{\pi}{2}$.
Four different values of explicity CPV parameters of ${\rm Im} \lambda_5=10^{-1}$ (upper left), ${\rm Im} \lambda_5=10^{-10}$ (upper right), $ {\rm Im} \lambda_5=10^{-15}$ (lower left), and $ {\rm Im} \lambda_5=10^{-20}$ (lower right) are taken.
}
\end{figure}

We shall solve for the relative CPV phase of $\theta$ in the ECPV scenario.
Different from the SCPV scenario, the input of ${\rm Im}\lambda_5$ also plays a role.
The related details are presented in Sec.~\ref{section:2HDMECPV}. 
In Fig.~\ref{fig:ECPV_theta}, we depicted $|s_{2\theta}|$ versus the CPV mixing angle of $\alpha_c\in (10^{-10}\,,1)$ with various inputs of ${\rm Im} \lambda_5$.
Indeed, for relatively sizeable values of ${\rm Im} \lambda_5 = 10^{-1}$ and ${\rm Im} \lambda_5 = 10^{-10}$, the solutions of $|s_{2\theta}|$ are plateaued when $\alpha_c$ drops to certain threshold.
Particularly for the ${\rm Im} \lambda_5 = 10^{-1}$ and $\Delta M=1\,\MeV$, the values of $s_{2\theta}$ is basically invariant with the varying $\alpha_c$.
When the ECPV parameters are suppressed to ${\rm Im} \lambda_5 = 10^{-15}$ and even ${\rm Im} \lambda_5 = 10^{-20}$, one finds that values of $|s_{2\theta}|$ are (almost) always decreasing with respect to the $\alpha_c$ inputs in the ranges of our consideration.
Together, they justify our approximations in Eqs.~\eqref{eq:ECPV_thetaApprox01} and \eqref{eq:ECPV_thetaApprox02}.

\subsection{The unitarity and stability constraints}

It is well-known that the perturbative unitarity and stability constraints to the Higgs potential should be imposed.
The unitarity bounds of the 2HDM were previously studied in Refs.~\cite{Arhrib:2000is,Kanemura:2015ska}.
The perturbative unitarity constraint means that the theory cannot be strongly coupled.
In practice, the necessary and sufficient condition of the tree-level unitarity bounds can be obtained by evaluating the eigenvalues of the $S$-matrices for the scattering processes of the scalar fields in the 2HDM~\cite{Arhrib:2000is, Kanemura:2015ska}.
Due to the Nambu-Goldstone theorem, the $S$-matrices can be expressed in terms of 2HDM quartic scalar self couplings $\lambda_i$.
The $S$-wave amplitude matrices are due to fourteen neutral, eight singly-charged, and three doubly-charged scalar channels in the 2HDM.
They read
\beqs
\beqn
\textrm{neutral}~~a_0^0&:& | H_i^+ H_i^- \rangle\,,\qquad | H_1^\pm H_2^\mp \rangle \,, \qquad  \frac{1}{\sqrt{2} } | A_i^0 A_i^0 \rangle\,, \qquad \frac{1}{\sqrt{2} } | H_i^0 H_i^0 \rangle\,, \non
&& | H_i^0 A_i^0 \rangle \,, \qquad | A_1^0 A_2^0 \rangle\,, \qquad | H_1^0 H_2^0 \rangle\,,\non
&&  | H_1^0 A_2^0 \rangle\,, \qquad  | H_2^0 A_1^0 \rangle\,,\\
\textrm{singly-charged}~~a_0^+&:& | H_i^+ A_i^0  \rangle\,, \qquad | H_i^+ H_i^0  \rangle\,, \non
&&| H_1^+ A_2^0  \rangle\,, \qquad | H_2^+ A_1^0  \rangle\,, \qquad | H_1^+ H_2^0  \rangle\,, \qquad | H_2^+ H_1^0  \rangle \,,\\
\textrm{doubly-charged}~~a_0^{++}&:& \frac{1}{\sqrt{2}} | H_i^\pm H_i^\pm \rangle \,,\qquad  | H_1^\pm H_2^\pm \rangle \,.
\eeqn
\eeqs
The $S$-wave amplitude matrices for three different channels are expressed as
\beqs
\beqn
&&a_0^0 = \frac{1}{16\pi} {\rm diag}(X_{4\times 4}\,, Y_{4\times 4}\,, Z_{3\times 3}\,, Z_{3\times 3})\,,\\
&& a_0^+ = \frac{1}{16\pi} {\rm diag} (Y_{4\times 4}\,, Z_{3\times 3}\,, \lambda_3 - \lambda_4)\,, \\
&& a_0^{++}=  \frac{1}{16\pi} Z_{3\times 3}\,.
\eeqn
\eeqs
The expressions for the submatrices of $(X_{4\times 4}\,, Y_{4\times 4}\,, Z_{3\times 3})$ are given as follows
\beqs
\beqn
X_{4\times 4}&=& \left(
\ba{cccc}
3 \lambda_1 & 2 \lambda_3 + \lambda_4 & 0 & 0 \\
2 \lambda_3 + \lambda_4 & 3 \lambda_2 & 0 & 0 \\
0  & 0 & \lambda_3 + 2\lambda_4 + 3 {\rm Re} \lambda_5   & 3 {\rm Im} \lambda_5   \\
0 & 0 & 3 {\rm Im} \lambda_5 & \lambda_3 + 2\lambda_4 - 3 {\rm Re} \lambda_5  \\  \ea \right) \,,\\
Y_{4\times 4}&=& \left(
\ba{cccc}
 \lambda_1 &  \lambda_4 & 0 & 0 \\
 \lambda_4 & \lambda_2 & 0 & 0 \\
0  & 0 & \lambda_3 + {\rm Re} \lambda_5   & {\rm Im}\lambda_5   \\
0 & 0 & {\rm Im}\lambda_5 & \lambda_3 - {\rm Re}\lambda_5  \\  \ea \right)  \,,\\
Z_{3\times 3}&=&  \left(
\ba{ccc}
 \lambda_1 &  {\rm Re}\lambda_5 + i {\rm Im} \lambda_5 & 0  \\
{\rm Re}\lambda_5 - i {\rm Im}\lambda_5 & \lambda_2 & 0  \\
0  & 0 & \lambda_3  + \lambda_4 \\  \ea \right)  \,.
\eeqn
\eeqs
The unitarity requires $|a_0^i|\leq 1$.

\begin{figure}[!tbp]
\centering
\includegraphics[width=0.6\textwidth]{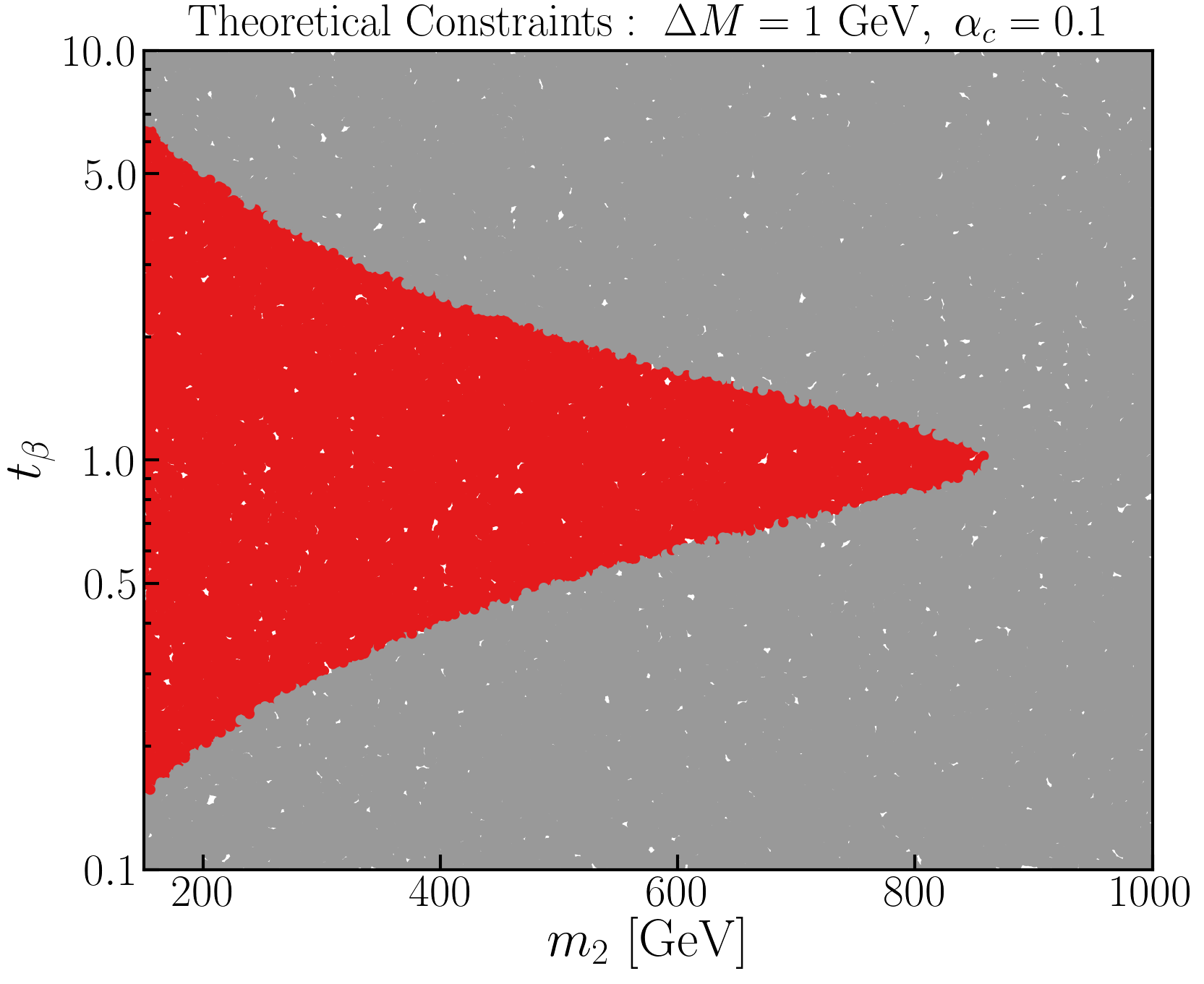}
\caption{The theoretical constraints in the $(m_2\,, t_\beta)$ plane with $\Delta M=1$ GeV, $\alpha_c = 0.1$, where the red (grey) regions are theoretically allowed (disfavored).
 }\label{fig:theoretical}
\end{figure}

The stability constraints require a positive 2HDM potential for large values of Higgs fields along all field space directions.
Collectively, they read
\beqn\label{eq:2HDM_stability}
&& \lambda_{1\,,2}>0\,, \qquad \lambda_3 > - \sqrt{\lambda_1 \lambda_2  }\,,\qquad \lambda_3 + \lambda_4 - | \lambda_5 | > - \sqrt{ \lambda_1 \lambda_2  } \,.   
\eeqn
In Fig.~\ref{fig:theoretical}, we present the theoretical constraints in the $(m_2\,, t_\beta)$ plane, with other parameters fixed to be $\Delta M = 1\,\GeV$ and $\alpha_c=0.1$.
The heavy neutral Higgs boson masses are found to be bounded from above, and large $t_\beta \gtrsim 5.0$ or small $t_\beta\lesssim 0.2$ are disfavored.


\section{Domain walls in the CPV 2HDM}
\label{section:2HDMDW}

In this section, we study the domain wall solutions in the 2HDM with the SCPV vacuum solution~\footnote{This was dubbed as the CP1 domain wall in the Ref.~\cite{Battye:2011jj}.}.
Such solutions arise from the CP transformations of $\Phi_j\to \Phi_j^*$ in the 2HDM.

\subsection{The domain wall solutions}

Under the discrete CP transformations of two Higgs doublets
\beqn\label{eq:CP1trans}
&& \Phi_1\to \Phi_1^* \,,\qquad  \Phi_2 \to \Phi_2^* \,,
\eeqn
one has $\Theta\to - \Theta$ for the background fields.
This means the CP transformation to two Higgs doublets is equivalent to a $\mathbb{Z}_2$ transformation to their relative phase.
Correspondingly, the CPC part of Eq.~\eqref{eq:2HDMpotential} is invariant, while the CPV terms in Eq.~\eqref{eq:2HDMpotential} is manifestly odd.
The vacuum manifold and the corresponding nontrivial homotopy group~\cite{Battye:2011jj,brawn_SymmetriesTopologicalDefects_2011} is
\beqn
&& \mM^{\rm SCPV} \simeq \mathbb{Z}_2 \otimes S^3\,,\qquad \Pi_0(  \mM^{\rm SCPV} ) \neq \mathbf{1}\,,
\eeqn
where one uses the fact that the CP symmetry is homeomorphic to the $\mathbb{Z}_2$ symmetry, and the  vacuum manifold of $\gSU(2)_L \times \gU(1)_Y / \gU(1)_{\rm em}$ is  homeomorphic to $S^3$.
Therefore, the SCPV part of the 2HDM potential leads to a domain wall solution.

\begin{figure}[!tbp]
    \centering
    \includegraphics[width=\textwidth]{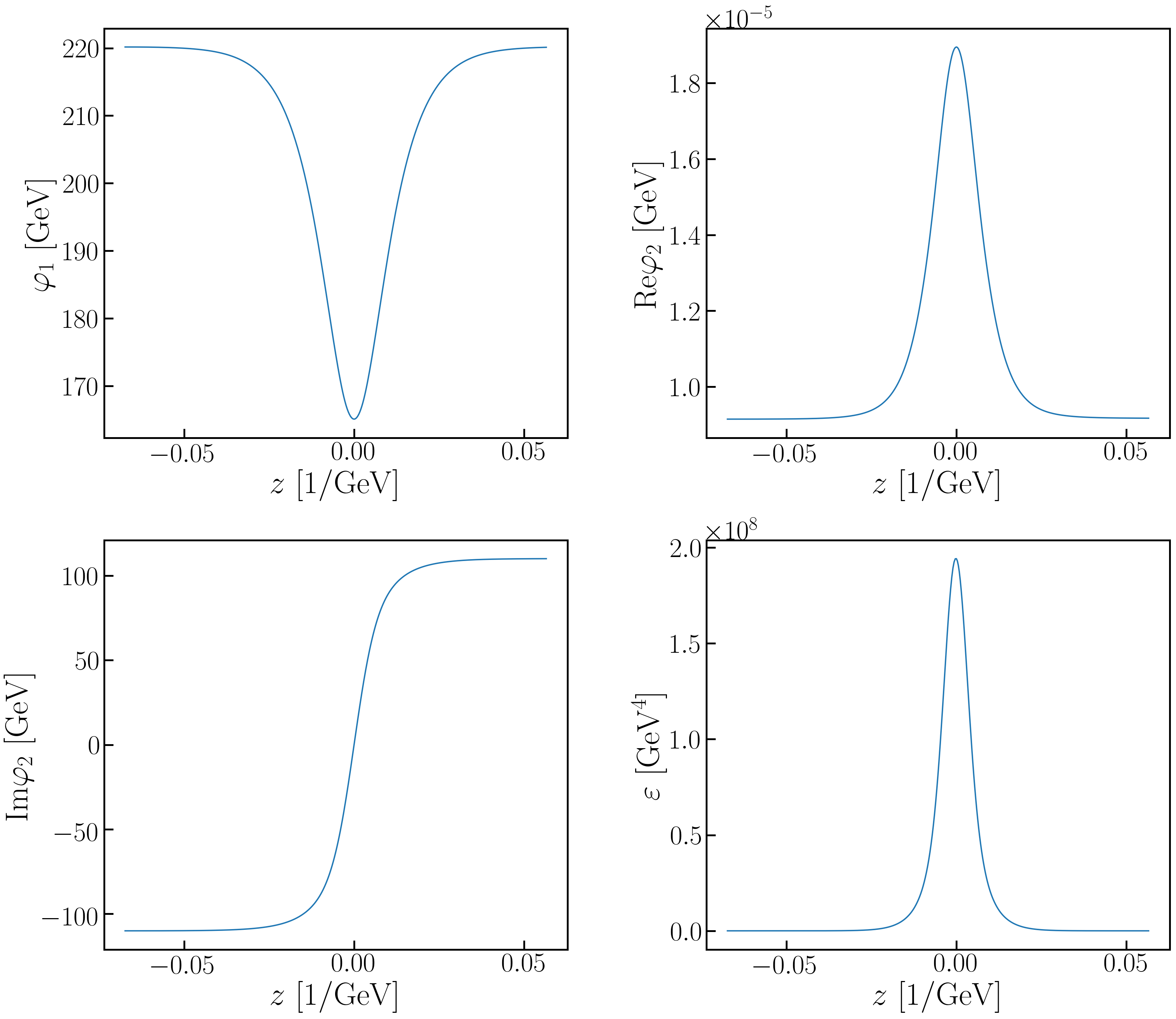}
\caption{
The profiles of the background fields and energy density across the domain wall for $m_2=400$ GeV, $\Delta M = 1$ GeV, $t_\beta = 0.5$, $\alpha = \beta - \pi/2$, $\alpha_c = 10^{-5}$.
}
    \label{fig:DW_profile}
\end{figure}

The domain wall solution is obtained in the `Euclidean basis' of
\begin{align}
  \vec{\phi} \equiv \left(\varphi_1, {\rm Re}\varphi_2, {\rm Im}\varphi_2\right) = \left(\varphi_1, \varphi_2 c_\Theta, \varphi_2  s_\Theta  \right).
\end{align}
Two domains correspond to $\vec{\phi}=(v_1,c_\theta v_2,\pm s_\theta v_2)$.
The energy density is given as follows:
\begin{subequations}\label{eqs:ECPV_density}
 \begin{align}
\mathcal{E}_{\rm total} &= \frac{1}{2}\left(\partial_z\vec{\phi}\right)^2 + V(\vec{\phi}),\\
   V(\vec\phi) &= V_{\rm CPC} + V_{\rm SCPV} + V_{\rm ECPV} + V_0,\\
V_{\rm CPC} &= \hf m_{11}^2 \varphi_1^2 + \hf m_{22}^2 \varphi_2^2 + \frac{\lambda_1}{8} \varphi_1^4 + \frac{\lambda_2}{8} \varphi_2^4 + \frac{1}{4} ( \lambda_{3} + \lambda_4 ) \varphi_1^2\varphi_2^2,\\
    V_{\rm SCPV} &= -{\rm Re}m_{12}^2\varphi_1\varphi_2 + \frac{1}{4}{\rm Re}\lambda_5\varphi_1^2 \varphi_2^2,\\
    V_{\rm ECPV} &= -\frac{1}{4}{\rm Im}\lambda_5 \varphi_1^2 \varphi_2^2 s_{2\Theta} \,,\\
    V_0 &=  \frac{1}{8} \left( \lambda_1 v_1^4 + \lambda_2 v_2^4  \right) + \frac{1}{4} \left( \lambda_{3}  + \lambda_4 + {\rm Re} \lambda_5 c_{ 2\theta }  \right) v_1^2 v_2^2.\label{eq:V0}
  \end{align}
\end{subequations}
where $V_0$ is a pure constant to make the potential of electroweak vacuum zero.
In our real calculation of the domain wall profile, $V_{\rm ECPV}$ should be taken into account, as it shifts the potential as well as the local minima positions.
However, when considering the case $|{\rm Im}\lambda_5|\ll 1$, we can safely ignore the ECPV part when solving the domain wall profile.
The tension of the domain wall is then the integral of the total energy density:
\begin{align}\label{eq:ECPV_tension}
  \sigma\simeq \int_{-\infty}^{\infty}dz\left[\frac{1}{2}\left(\partial_z\vec{\phi}\right)^2 + V(\vec{\phi})\right]\,.
\end{align}
The corresponding domain wall profile is obtained by solving the equations of motion (EOM) of $\vec{\phi}$:
\begin{align}
  \frac{d^2}{dz^2}\vec{\phi} = \vec{\nabla}_{\phi}V(\vec{\phi})\,,
\end{align}
with the boundary conditions being
\begin{align}
  \vec{\phi}(z=\mp \infty) &= (v_1\,,  v_2 c_\theta\,, \mp v_2 s_\theta )\,.
  \end{align}
The EOM is solved using the path deformation algorithm~\footnote{The corresponding code can be found in \url{https://github.com/ycwu1030/BSM_Soliton}.}~\cite{Wainwright:2011kj,Chen:2020wvu}.
In Fig.~\ref{fig:DW_profile}, we display a sample of the domain wall profiles and the energy density of $\mathcal{E}(z)$.
Since we expect that the relative CPV phase to be $\theta \to \pm \pi/2$, thus, the imaginary component of $\varphi_2$ takes much larger value comparing to the real component.
In our numerical estimation, we find that the domain wall tensions from Eq.~\eqref{eq:ECPV_tension} are typically $\sim \mO(10^6)\, \GeV^3$.
One can quickly find this result from the energy density plot of our sample in Fig.~\ref{fig:DW_profile}, where the local energy density is roughly $\mathcal{E} \sim \mO(10^8)\,\GeV^4$ and the domain wall width is around $\sim \mO(0.01)\,\GeV^{-1}$.

\subsection{The biased term from the ECPV}

The ECPV component of the potential leads to additional contribution to the energy density as below
\beqn
V_{ \rm ECPV  }(z) &=&- \frac{1}{4} {\rm Im} \lambda_5   \varphi_1^2 \varphi_2^2 s_{2 \Theta} \,,
\eeqn
which becomes the biased term for the SCPV domain walls.
After the EWSB, the corresponding biased term reads
\beqn
\Delta V&=& \Big| V_{ \rm ECPV  }(z= + \infty) - V_{ \rm ECPV  }(z= - \infty)   \Big|  \non
&=&  \hf \Big|{\rm Im} \lambda_5 s_{2 \theta } \Big|  v_1^2 v_2^2 \,,
\eeqn
with the relative CPV phase of $\theta$ being solved from Eq.~\eqref{eq:ECPV_theta_ExactSol}.

\subsection{The cosmological constraints}

The observation of the CMBR leads to the following condition to the domain wall tension
\beqn\label{eq:ZKO_bound}
&&\sigma \lesssim \mO(1)\,\MeV^3\,,
\eeqn
which was known as the Zel'dovich-Kobzarev-Okun bound.
Our numerical solutions found that the domain wall solutions in the SCPV case generally lead to tensions of $\sigma \sim \mO(10^6)\,\GeV^3$, which is $\sim 10^{15}$ times above the Zel'dovich-Kobzarev-Okun bound.
Therefore, the biased term from the ECPV is necessary for the domain wall collapse.

To have the large scale domain wall to form, it is known that the energy difference between two vacua should be sufficiently small:
\beqn\label{eq:DeltaV_upper}
&& \frac{\Delta V}{ | V_0| } < \log \frac{1-p_c}{p_c} = 0.795\,,
\eeqn
with $|V_0|$ representing the height of the potential barrier between two minima in Eq.~\eqref{eq:V0}, and the critical value of $p_c=0.311$ predicted from the percolation theory~\cite{Stauffer:1978kr}.

\begin{figure}[htb]
\centering
\includegraphics[height=5cm]{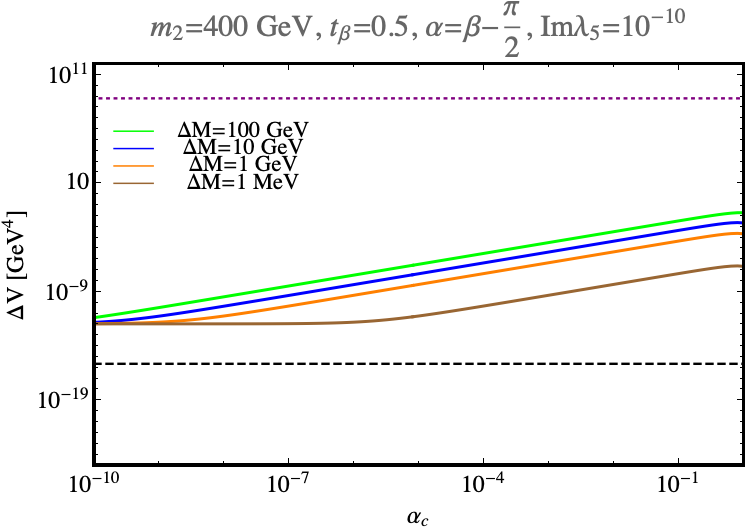}
\includegraphics[height=5cm]{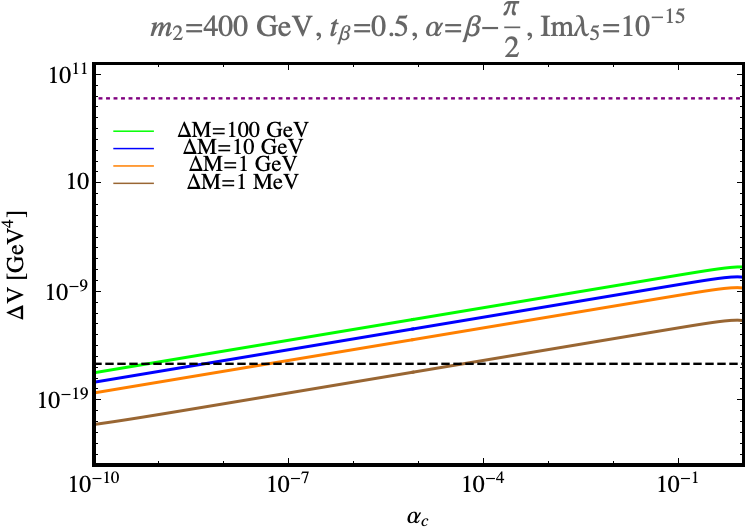}\\
\includegraphics[height=5cm]{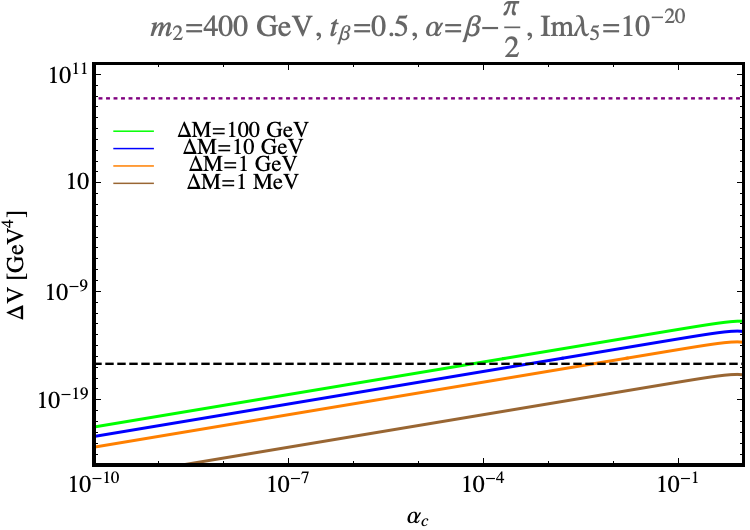}
\includegraphics[height=5cm]{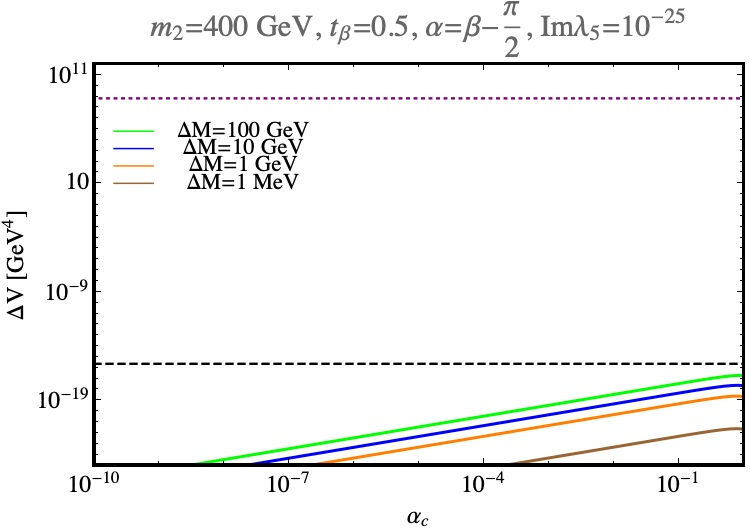}
\caption{\label{fig:DeltaV_alphac}
The $\Delta V$ versus the $\alpha_c$ with various inputs of: ${\rm Im} \lambda_5=10^{-10}$ (upper-left), ${\rm Im} \lambda_5=10^{-15}$ (upper-right),  ${\rm Im}  \lambda_5=10^{-20}$ (lower-left), and ${\rm Im}  \lambda_5=10^{-25}$ (lower-right).
In each plot, four different mass splittings of $\Delta M = (100\,\GeV\,, 10\,\GeV\,, 1\,\GeV\,, 1\,\MeV )$ are presented.
Other input parameters are $m_2=400\,\GeV$ and $t_\beta =0.5$.
The dotted lines represent the upper bound to the domain wall formation in Eq.~\eqref{eq:DeltaV_upper}, and the dashed lines represent the lower bound to the domain wall collapse in Eq.~\eqref{eq:DeltaV_lower}.
}
\end{figure}

Lower bounds can be also obtained to the energy difference.
The domain wall cannot exist too long to spoil the known constraints from the BBN~\cite{Kawasaki:2004yh,Kawasaki:2004qu,Saikawa:2017hiv}.
This leads to an lower bound to the energy difference
\beqn\label{eq:DeltaV_lower}
&& \Delta V^{1/4} \gtrsim 5.07 \times 10^{-4}\,{\rm GeV}\, C_{\rm ann}^{1/4}  \mA^{1/4} \hat \sigma^{1/4}\,,
\eeqn
with $\mA\sim 0.8\pm 0.1$~\cite{Hiramatsu:2013qaa}, and $C_{\rm ann} =2$ for the $\mathbb{Z}_2$ symmetry.
$\hat \sigma \equiv \sigma / (1\,\TeV )^3$ denotes the dimensionless domain wall tension.
Assuming that the domain wall collapse occurred during the radiation dominated era, the corresponding temperature is given by
\beqn\label{eq:Tann}
T_{\rm ann}&=& 3.41\times 10^{-2}\,{\rm GeV}\, C_{\rm ann}^{-1/2}   \mA^{-1/2}  \Big(  \frac{ g_*(T_{\rm ann}) }{10}  \Big)^{-1/4}  \hat\sigma^{-1/2} \Delta\hat V^{1/2}  \,,
\eeqn
with $g_*(T_{\rm ann})$ counting the relativistic degrees of freedom at the annihilation temperature, and $\Delta\hat V \equiv \Delta V /(1\,\MeV)^4$.

In Fig.~\ref{fig:DeltaV_alphac}, we present the size of the biased term $\Delta V$ versus the CPV mixing angle $\alpha_c\in (10^{-10}\,, 1)$, for different mass splittings of $\Delta M=(100\,\GeV\,, 10\,\GeV\,, 1\,\GeV\,, 1\,\MeV)$ between two neutral heavy Higgs bosons.
The upper bound from Eq.~\eqref{eq:DeltaV_upper} and lower bound from Eq.~\eqref{eq:DeltaV_lower} are presented in terms of dotted (in purple) and dashed (in black) lines, respectively.
It turns out that the upper bound can be always satisfied with the parameter choices of $\alpha_c$ in our considerations.
Meanwhile, the lower bounds to the domain wall collapse can become sensitive to the CPV mixing angle of $\alpha_c$ only when the explicit CPV parameter of ${\rm Im} \lambda_5$ is sufficiently small.
This can be expected from our previous discussions about the $t_\theta$ dependences on the physical inputs of $\alpha_c$ and ${\rm Im }\lambda_5$.
For the ${\rm Im} \lambda_5=10^{-10}$ case, one also finds that $\Delta V$ becomes plateaued similar to the corresponding $|s_{2\theta} |$ plot in Fig.~\ref{fig:ECPV_theta}.
On the other hand, too small $\Delta M$ and/or ${\rm Im}\lambda_5$ values are excluded by the BBN lower limit as shown in the lower panels of Fig.~\ref{fig:DeltaV_alphac}.

\subsection{The GW signals}

The collapsing domain walls can lead to GW signals~\cite{Gleiser:1998na,Hiramatsu:2010yz,Kawasaki:2011vv,Hiramatsu:2013qaa,Saikawa:2017hiv}, while we find that such signals arising from the CPV 2HDM are impossible to be probed in any of the future satellite observations.
We shall briefly discuss the signal estimations below.

The peak frequency of the GWs at the annihilation time of domain walls is proportional to the annihilation temperature $T_{\rm ann}$ in Eq.~\eqref{eq:Tann}, and is given by
\beqn\label{eq:fpeak}
f_{\rm peak}& \simeq & 1.1\times 10^{-7}\,{\rm Hz}\, \Big(  \frac{ g_*( T_{\rm ann} ) }{10}  \Big)^{1/2}  \Big(  \frac{ g_{*s}( T_{\rm ann} ) }{10}  \Big)^{-1/3} \Big(  \frac{ T_{\rm ann} }{1\,\GeV }  \Big)\non
&\simeq&  3.75\times 10^{-9}\,{\rm Hz}\,  \Big(  \frac{ g_*( T_{\rm ann} ) }{10}  \Big)^{1/4} \Big(  \frac{ g_{*s}( T_{\rm ann} ) }{10}  \Big)^{-1/3} C_{\rm ann}^{-1/2} \mA^{-1/2} \hat\sigma^{-1/2}   \Delta \hat V^{1/2}    \,.
\eeqn
Here, $g_*(T_{\rm ann})$ and $g_{*s}(T_{\rm ann})$ count the relativistic degrees of freedom contributing to the energy density and the entropy density.
They are both $10.75$ for $1\,\MeV\lesssim T_{\rm ann} \lesssim 100\,\MeV$.
For GWs with peak frequencies in the range of $\mO(10^{-4}) - \mO(10^{-1})\,{\rm Hz}$, they may be probed by the future satellite-based interferometers, such as the LISA~\cite{AmaroSeoane:2012km,AmaroSeoane:2012je}, Taiji~\cite{Guo:2018npi}, and Tianqin~\cite{Luo:2015ght} programs.
The GWs with very small peak frequencies of few nano Hz may be probed at the future radio telescope of SKA~\cite{Janssen:2014dka} and the  DECIGO~\cite{Kawamura:2011zz} with the latter having wider range of typical frequencies of $\sim \mO(0.1) - \mO(10) $ Hz.
With the lower limit of the $\Delta V$ in Eq.~\eqref{eq:DeltaV_lower}, there is a lower limit to the peak frequency of
\beqn\label{eq:fpeak_lower}
f_{\rm peak}&\gtrsim & 0.94 \times 10^{-9}\,{\rm Hz}\Big(  \frac{ g_*( T_{\rm ann} ) }{10}  \Big)^{1/4} \Big(  \frac{ g_{*s}( T_{\rm ann} ) }{10}  \Big)^{-1/3}  \,.
\eeqn
Thus, the peak frequencies of the GW signals are expected to be higher than order of several nano Hz.

The peak energy density spectrum of the GW is
\beqn
\Omega_{\rm GW}^{\rm peak} h^2(t_0)&=& 7.2\times 10^{-18} \, \tilde \epsilon_{\rm GW} \mA^2 \Big( \frac{ g_{*s}(T_{\rm ann})  }{10}   \Big)^{-4/3}   \Big(  \frac{ T_{\rm ann} }{ 10^{-2}\,\GeV } \Big)^{-4} \hat \sigma^2 \,,
\eeqn
with $\tilde \epsilon_{\rm GW}\simeq 0.7\pm 0.4$ in the scaling regime~\cite{Hiramatsu:2013qaa}.
By using the annihilation temperature in Eq.~\eqref{eq:Tann}, the peak energy density spectrum becomes
\beqn
\Omega_{\rm GW}^{\rm peak} h^2(t_0)&=& 5.3\times 10^{-20}\, \tilde \epsilon_{\rm GW} \mA^4 C_{\rm ann}^2 \Big( \frac{ g_{*s}(T_{\rm ann})  }{10}   \Big)^{-4/3}   \Big(  \frac{ g_{*} (T_{\rm ann})  }{10}  \Big) \hat \sigma^4 \Delta \hat V^{-2}  \,.
\eeqn
By taking the lower limit of the $\Delta V$ in Eq.~\eqref{eq:DeltaV_lower} into account, we find an upper limit to the peak energy density spectrum as
\beqn
\Omega_{\rm GW}^{\rm peak} h^2(t_0)&\lesssim & 1.36 \times 10^{-17} \tilde \epsilon_{\rm GW} \mA^2 \Big( \frac{ g_{*s}(T_{\rm ann})  }{10}   \Big)^{-4/3}   \Big(  \frac{ g_{*} (T_{\rm ann})  }{10}  \Big) \hat \sigma^2\,.
\eeqn
With the lower limit of the peak frequencies $\gtrsim 10^{-9}\,$Hz in Eq.~\eqref{eq:fpeak_lower}, and the peak energy density to be less than $\sim \mO(10^{-23})$, by using the facts that $\sigma \sim \mO(10^6)\,\GeV^3$ through the numerical results.
The related GW signals from the domain wall collapsing are roughly $\sim \mO(10^{-8})$ below the search sensitivities of SKA.


\section{The EDM measurements}
\label{section:EDM}

Obviously, the EDM measurements provide us direct constraints to the size of the CPV.
They can provide exclusive bound on the size of CPV in 2HDM together with the cosmological constraint on domain wall.
In this work, we focus on the evaluations of the eEDM.
We estimate the eEDM from the CPV 2HDM with the domain wall solutions, together with the ECPV parameters.
The latest upper bound of the electric dipole moment of the electron from the ACME-II~\cite{Andreev:2018ayy} reads
\beqn\label{eq:ACME_II}
\textrm{ACME-II}&:& |\frac{d_e}{e} | \leq 1.1 \times 10^{-29}\,  {\rm cm}\,,
\eeqn
and the future projected precision from the ACME-III reads~\footnote{This projection can be found in \url{http://www.electronedm.org/}. }
\beqn\label{eq:ACME_III}
\textrm{ACME-III}&:& | \frac{d_e}{e} | \lesssim 1.0 \times 10^{-30}\,  {\rm cm}\,.
\eeqn
The SM estimations of the eEDM were of size $|d_e/e | \sim 10^{-44}\, {\rm cm}$ from the four-loop contributions of the CKM phase, and $|d_e/e | \sim 10^{-38} - 10^{-39}\, {\rm cm} $ by considering the CP-odd electron-nucleon interaction~\cite{Hoogeveen:1990cb,Pospelov:1991zt,Pospelov:2013sca,Yamaguchi:2020dsy}.

\begin{figure}[htb]
\centering
\includegraphics[height=4cm]{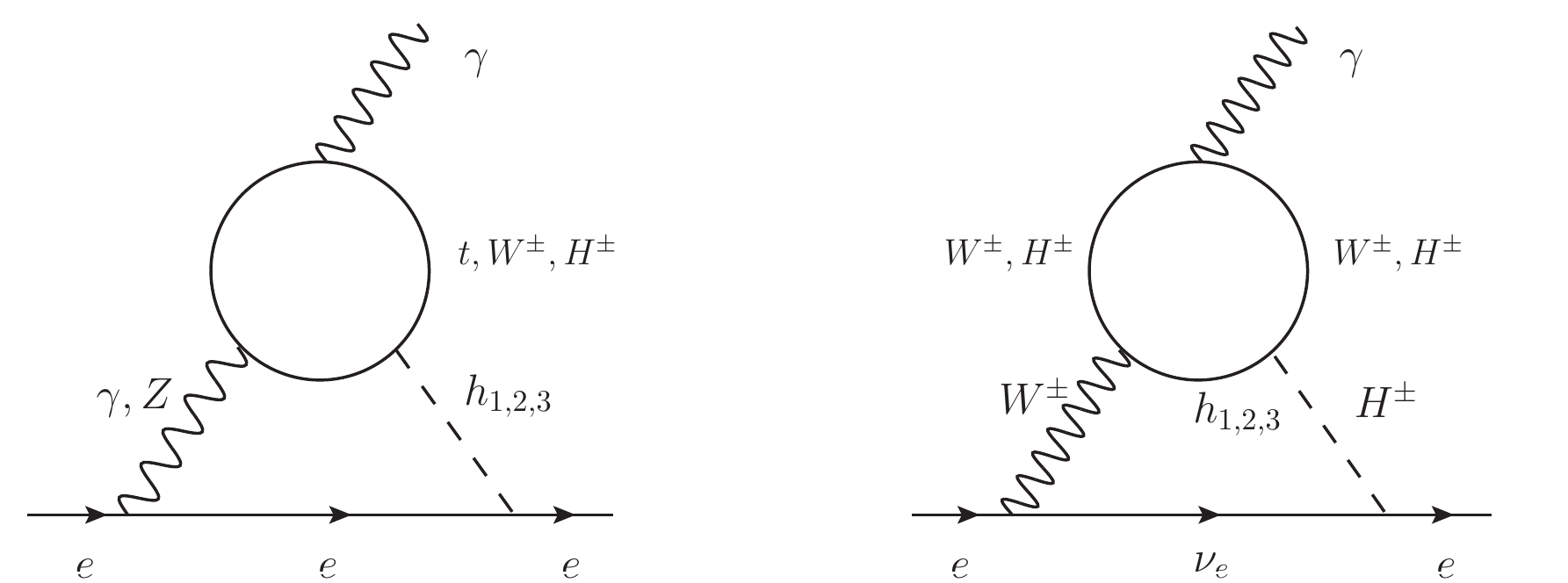}
\caption{\label{fig:BZ}
The two-loop Barr-Zee diagrams for the $h_i \gamma\gamma$, $h_i \gamma Z$, and $H^\pm W^\mp \gamma$ operators of the eEDM contributions.}
\end{figure}

The dominant contributions to the eEDMs come from the two-loop Barr-Zee diagrams~\cite{Barr:1990vd} in the CPV 2HDM.
There are three types of dimension-five operators involved: (i) the $h_i \gamma\gamma$ operator, (ii) the $h_i \gamma Z$ operator, and (iii) the $H^\pm W^\mp \gamma $ operator.
Expressed in terms of the Wilson coefficient of these operators, we summarize the total contributions as follows
\beqn\label{eq:eEDM_Wilson}
\delta_e&=&  (\delta_e )_t^{h_i \gamma\gamma } + (\delta_e )_W^{h_i \gamma\gamma } + (\delta_e )_{H^\pm}^{h_i \gamma\gamma }\non
&+& (\delta_e )_t^{h_i \gamma Z } + (\delta_e )_W^{h_i \gamma Z } + (\delta_e )_{H^\pm }^{h_i \gamma Z } + (\delta_e )_{h_i}^{H^\pm W^\mp \gamma} \,,
\eeqn
and the corresponding diagrams are depicted in Fig.~\ref{fig:BZ}.
Explicit expressions of these Wilson coefficients can be found in the appendices of Refs.~\cite{Inoue:2014nva,Cheung:2020ugr}.
By combining the total contributions in Eq.~\eqref{eq:eEDM_Wilson}, the eEDM is obtained by
\beqn\label{eq:eEDM}
\frac{d_e}{ e}&=& \frac{2 m_e }{ v^2 } \delta_e \,.
\eeqn
The Wilson coefficients are generally related to the normalized scalar or pseudoscalar Yukawa couplings as
\beqn
&& (\delta_e )_t \propto c_{\ell \,,i} \tilde c_{u\,,i}/ \tilde c_{\ell \,,i}  c_{t\,,i} \,,\qquad (\delta_e )_W \propto a_i \tilde c_{\ell\,,i}\,.
\eeqn
The pseudoscalar couplings are all proportional to the CPV mixing angles as $\tilde c_{f\,,i} \propto \alpha_c$, which can be found in Tab.~\ref{tab:Hcouplings} and the relation of Eq.~\eqref{eq:alphab_Approx}.
One can expect that future improvements of the eEDM precisions by an order of magnitude will further constrain the CPV mixing angle by an order of magnitude.

\begin{figure}[htb]
\centering
\includegraphics[width=0.45\textwidth]{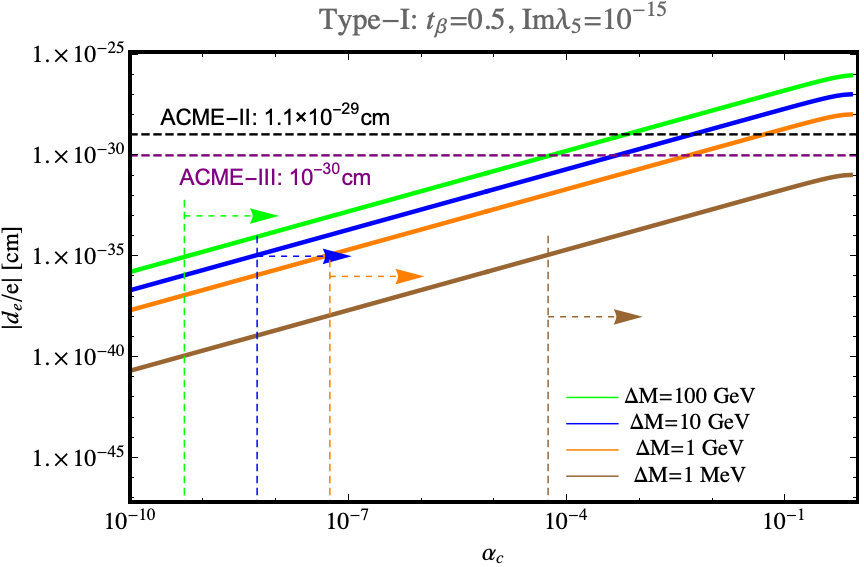}
\includegraphics[width=0.45\textwidth]{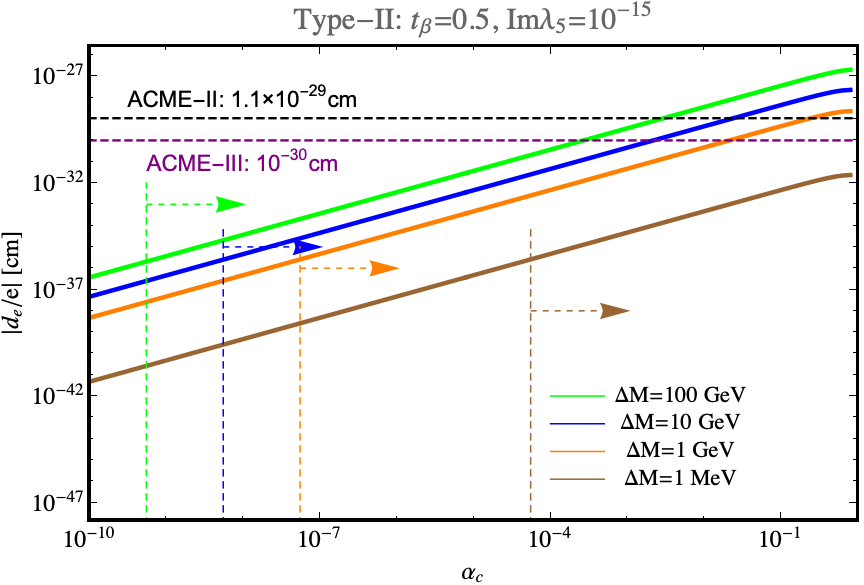}\\
\includegraphics[width=0.45\textwidth]{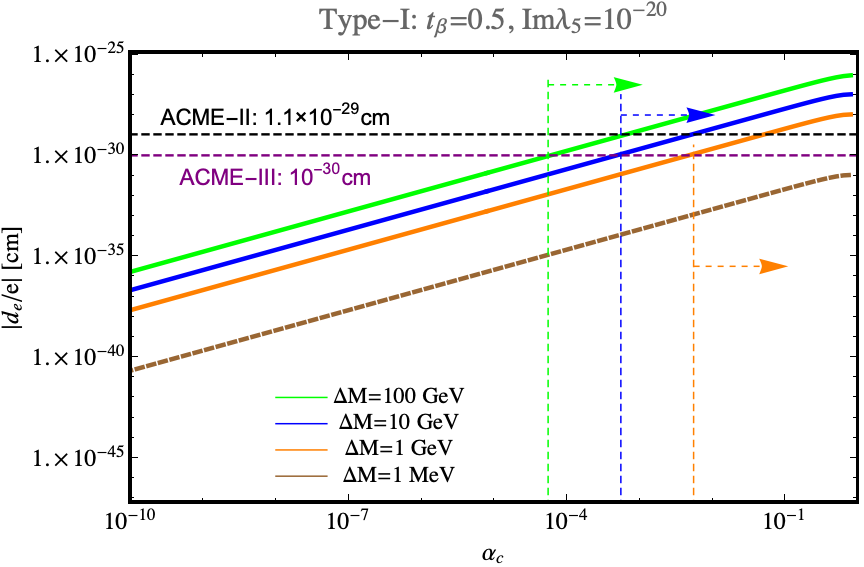}
\includegraphics[width=0.45\textwidth]{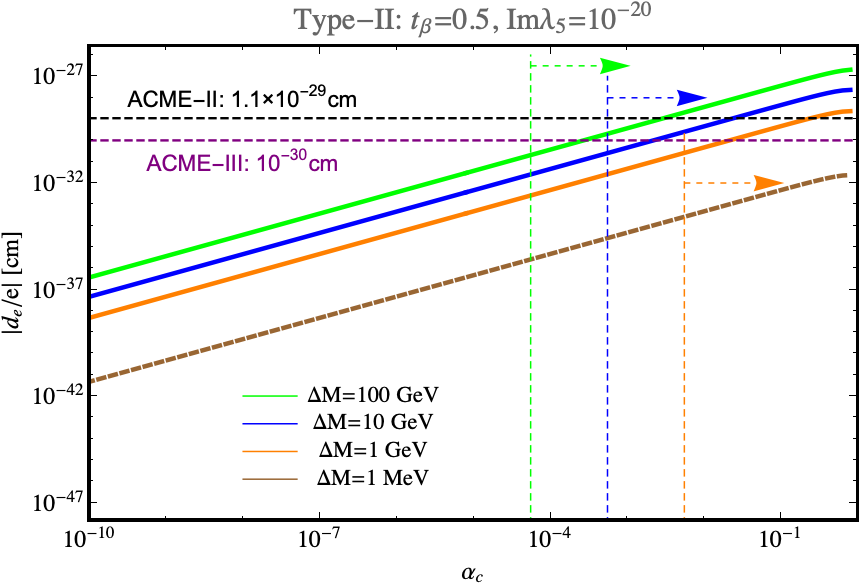}
\caption{\label{fig:eEDM}
The eEDM predictions versus the CPV mixing angle $\alpha_c$ in the range of $\alpha_c\in (10^{-10}\,,1)$.
The lower limits to $\alpha_c$ are shown in arrows for the ${\rm Im} \lambda_5 =10^{-15}$ (upper panels) and ${\rm Im} \lambda_5 =10^{-20}$ (lower panels) cases from the BBN bound to the biased term, as obtained from Eq.~\eqref{eq:DeltaV_lower}.
For each case, we display the eEDM predictions for different mass splittings of $\Delta M =(100\,\GeV\,, 10\,\GeV\,, 1\,\GeV\,, 1\,\MeV)$.
}
\end{figure}

We display the eEDM predictions of $|d_e /e |$ versus the CPV mixing angle of $\alpha_c$ in Fig.~\ref{fig:eEDM}.
The latest upper bound to the eEDM from the ACME-II from Eq.~\eqref{eq:ACME_II} and the future projected upper bound from the ACME-III in Eq.~\eqref{eq:ACME_III} are displayed in horizontal dashed lines.
The evaluations of the eEDM depend on the mixing angles of $(\alpha\,,\beta\,, \alpha_b\,,\alpha_c)$.
By using the constraint for masses and mixing angles in Eq.~\eqref{eq:mass_constraint}, the mass splitting of $\Delta M $ can also play a role in the size of the eEDM predictions.
With a fixed input of $\alpha_c$, one generally has a suppressed value of $\alpha_b$ with smaller inputs of $\Delta M$, as was displayed in Fig.~\ref{fig:alphab}.
Consequently, the eEDM predictions will be suppressed as well.
This was previously discussed in Ref.~\cite{Bian:2016zba}.
The evaluations of the eEDMs are independent of the size of ECPV parameter ${\rm Im} \lambda_5$, as one can visualize between two upper panels and two lower panels in Fig.~\ref{fig:eEDM}.
Meanwhile, different inputs of ${\rm Im} \lambda_5$ lead to different cosmological constraints to the $\alpha_c$ through Eq.~\eqref{eq:DeltaV_lower}.
Explicitly, these lower bounds to $\alpha_c$ are denoted by dashed vertical lines with arrows in each plot.
For the ${\rm Im} \lambda_5=10^{-15}$ case, there are lower limits to the CPV mixing angle of $\alpha_c$ for all four $\Delta M$ inputs.
However, when such lower limits of $\alpha_c$ are saturated, the corresponding eEDM predictions are $|d_e/e| \sim \mO(10^{-35})\,{\rm cm}$, which is another five orders of magnitude below the future precision aimed at the ACME-III.
When one further reduces the ECPV parameter to ${\rm Im} \lambda_5=10^{-20}$, the constraint of Eq.~\eqref{eq:DeltaV_lower} has already ruled out the situation with small mass splitting of $\Delta M =1\,\MeV$.
This can be also observed in the lower-left panel of Fig.~\ref{fig:DeltaV_alphac}.
Thus, we denote the $\Delta M =1\,\MeV$ with ${\rm Im} \lambda_5=10^{-20}$ cases by dashed lines, indicating that their exclusion from the BBN constraint.
We also checked that for the ${\rm Im} \lambda_5=10^{-20}$ case, the lower limits of $\alpha_c$ from Eq.~\eqref{eq:DeltaV_lower} lead to the eEDM predictions of $|d_e /e | \sim \mO(10^{-30})\, {\rm cm}$.
Therefore, the situation with very tiny ECPV parameter of  ${\rm Im} \lambda_5=10^{-20}$ is expected to be confirmed or excluded with the joint BBN constraints and the improved experimental precision of the eEDM from the future ACME-III.

\begin{figure}[!tbp]
    \centering
    \includegraphics[width=0.45\textwidth]{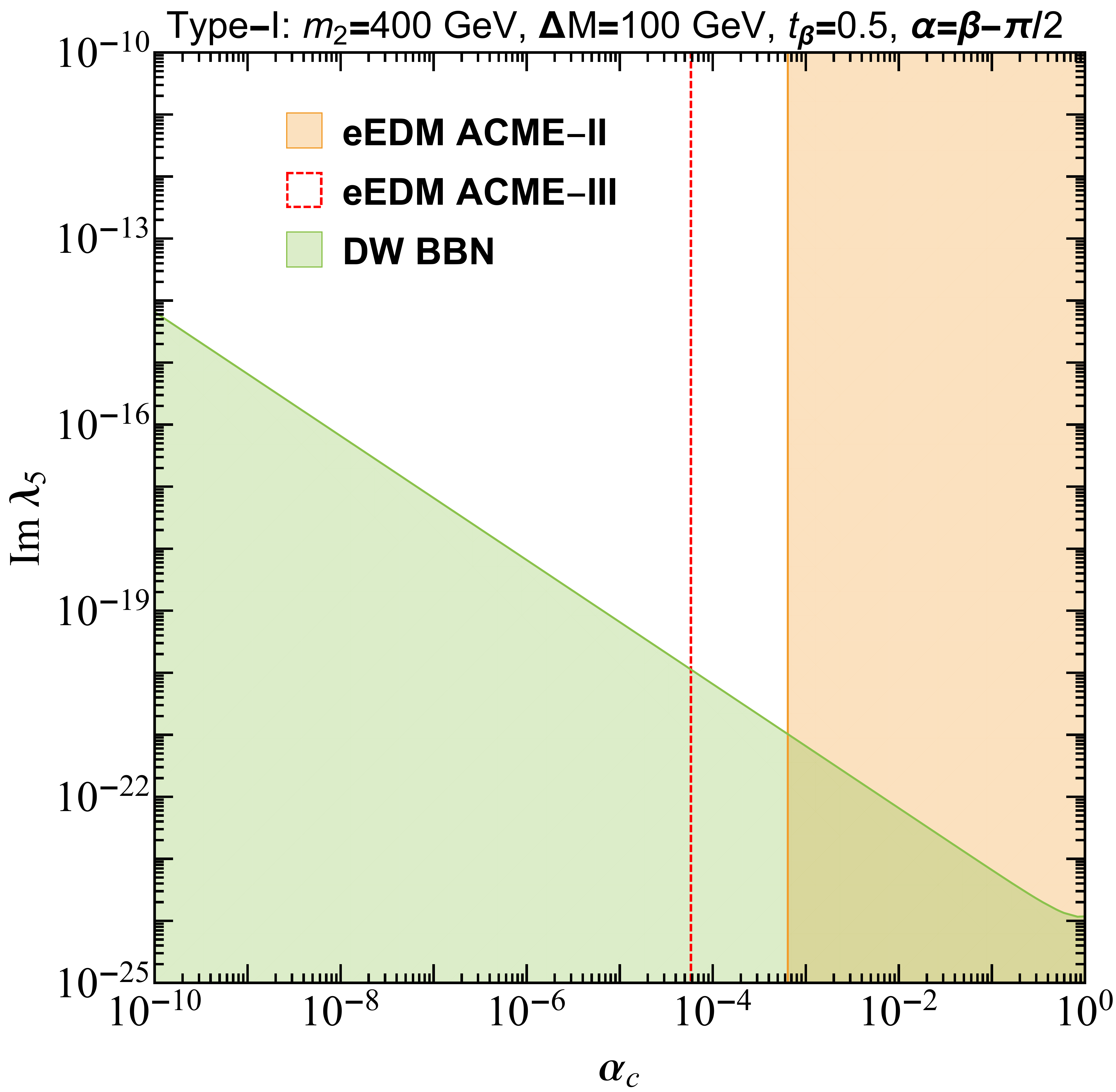}
    \includegraphics[width=0.45\textwidth]{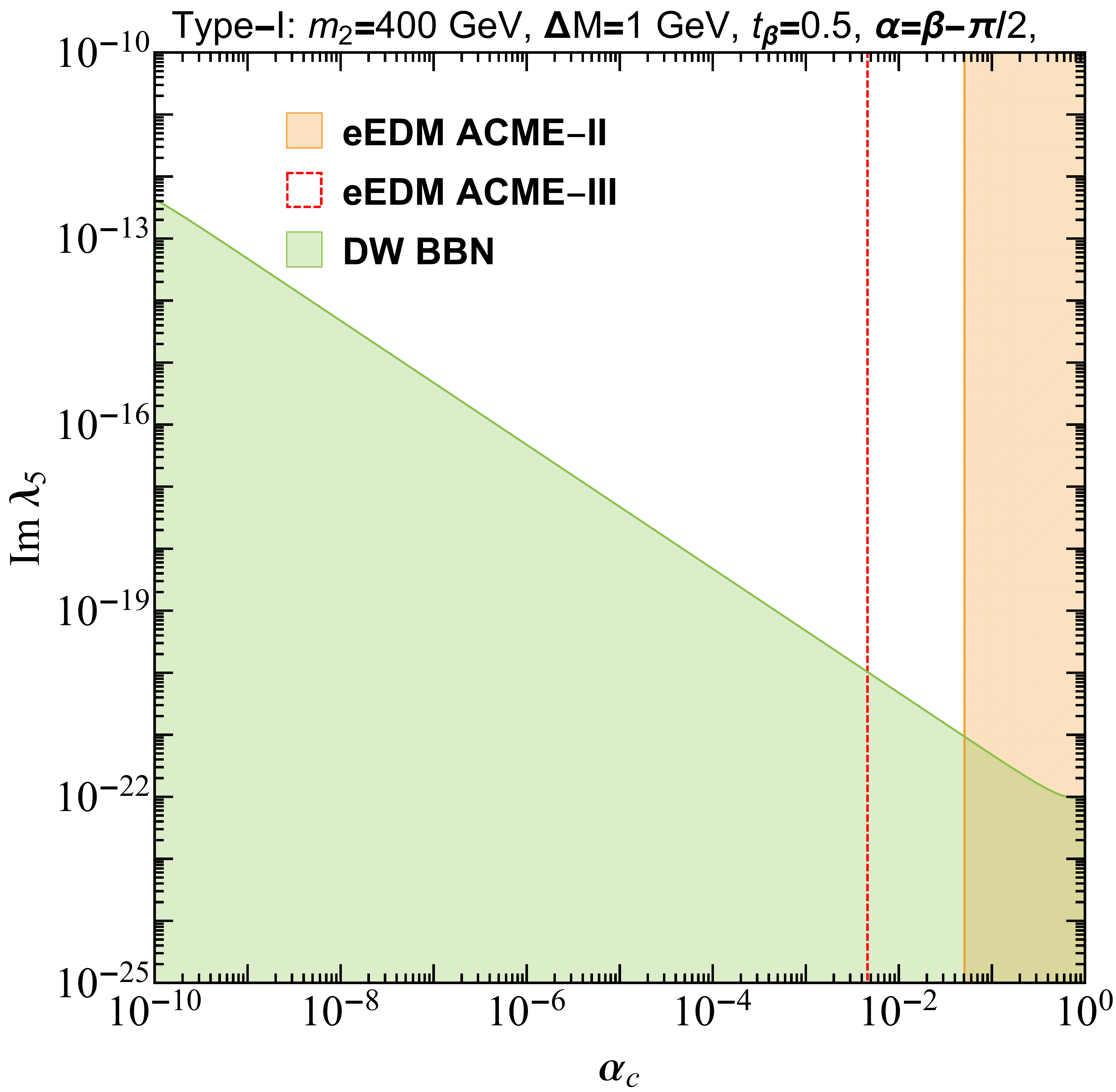}\\
    \includegraphics[width=0.45\textwidth]{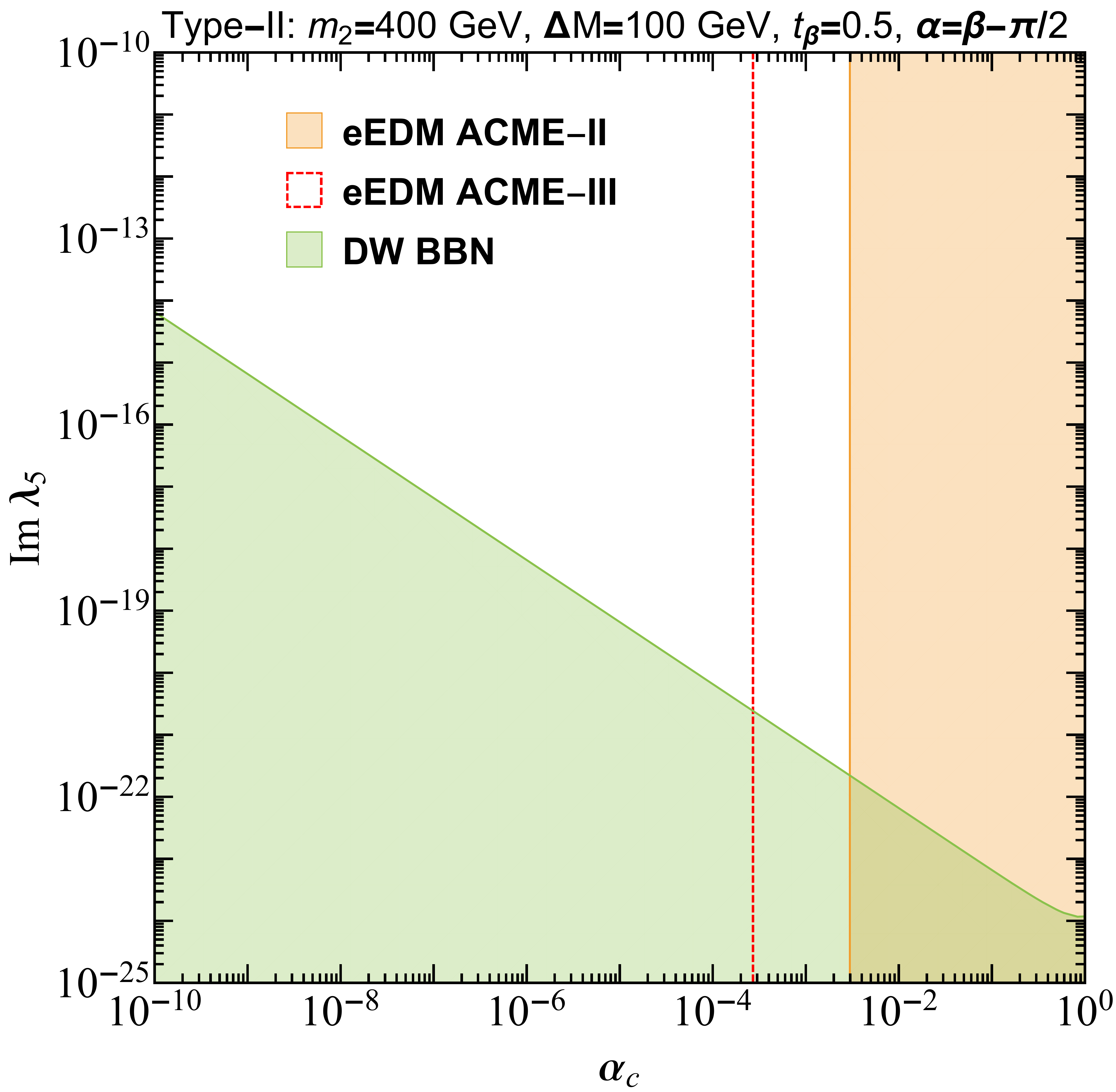}
    \includegraphics[width=0.45\textwidth]{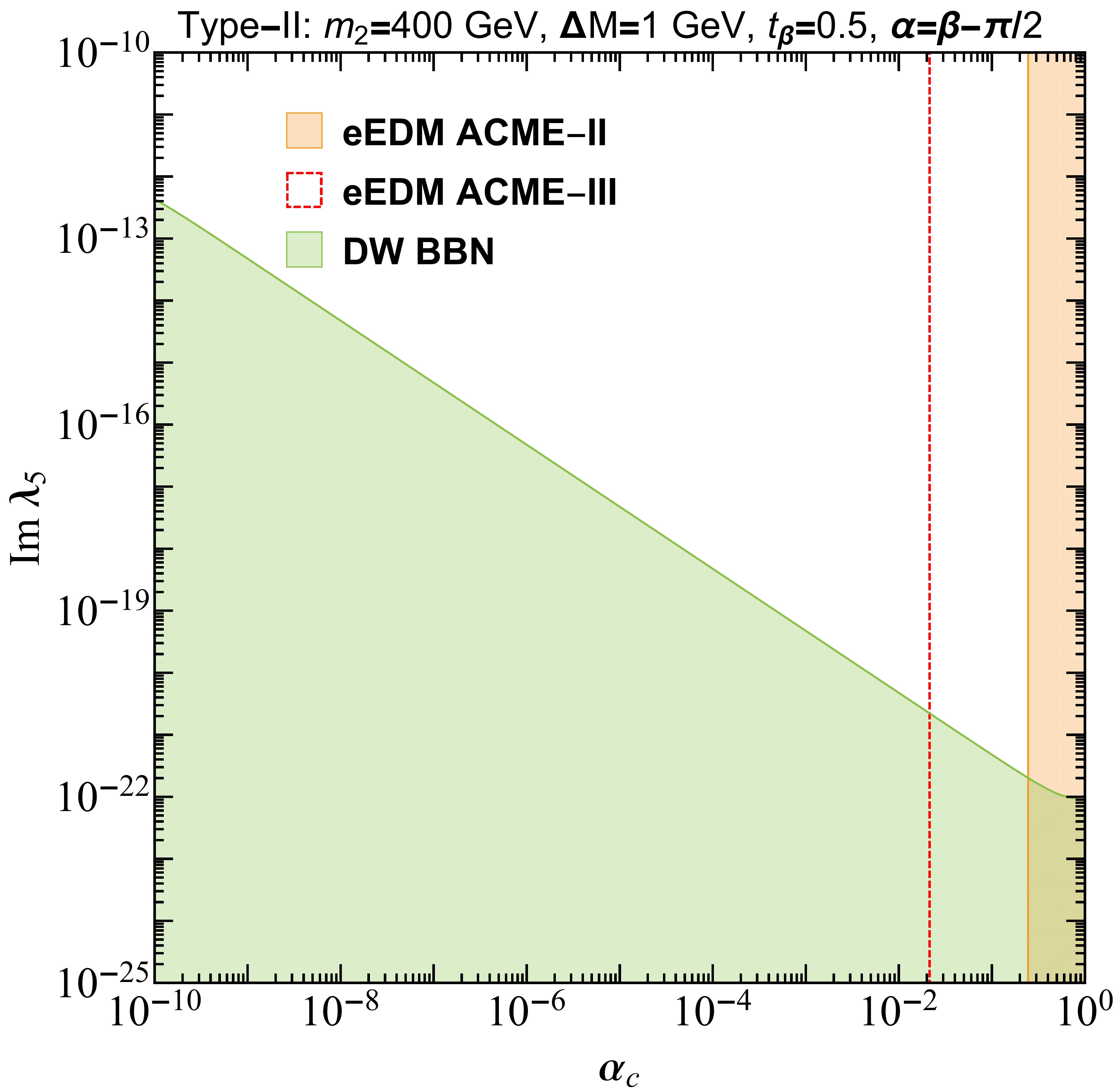}\\
    \caption{
    The joint BBN and eEDM constraints in the  $(\alpha_c\,, {\rm Im}\lambda_5)$ plane.
    The colored regions are excluded by corresponding constraints.
    The dashed lines correspond to the future improvements to the eEDM measurements from the ACME-III.
    }
    \label{fig:eEDM-DW-constraints}
\end{figure}

In Fig.~\ref{fig:eEDM-DW-constraints}, we further present the joint BBN constraint from Eq.~\eqref{eq:DeltaV_lower} and the eEDM measurements from the current ACME-II limits~\eqref{eq:ACME_II} (orange shaded regions) and the future ACME-III projections~\eqref{eq:ACME_III} (vertical dashed lines).
Two different mass splittings of $\Delta M=100\,\GeV$ (left panels) and $\Delta M=1\,\GeV$ (right panels) are displayed. With the future improvements of the eEDM precision by an order of magnitude, the corresponding constraints to the CPV mixing of $\alpha_c$ will be improved by an order of magnitude accordingly.
For a relatively large mass splittings of $\Delta M=100\,\GeV$, the upper limits to $\alpha_c$ from the future eEDM measurements can be as small as $\sim \mO(10^{-4})$.
While for a suppressed mass splittings of $\Delta M=1\,\GeV$, the upper limits to $\alpha_c$ become $\sim \mO(10^{-2})$.
This is due to the relation between $|\alpha_b |$ versus $\alpha_c$ as given in Eq.~\eqref{eq:alphab_Approx} and Fig.~\ref{fig:alphab}.

Furthermore, with the BBN constraints to the biased domain wall term, the sizes of the ECPV parameter ${\rm Im} \lambda_5$ are constrained with various inputs of the physical CPV mixing angle $\alpha_c$.
Such constraints are becoming more stringent with smaller inputs of $\alpha_c$. 
For a fixed value of ${\rm Im}\lambda_5$, a lower limit to $\alpha_c$ is given in the light green region.
We find that ${\rm Im}\lambda_5$ should be $\gtrsim \mO( 10^{-14})$ ($\gtrsim \mO( 10^{-24}$) with the CPV mixing angles of $\alpha_c\sim \mO(10^{-10})$ ($\alpha_c\sim \mO(1)$).
Joined with the current upper limits to the eEDM, we find that the incredibly tiny ECPV parameter ${\rm Im} \lambda_5$ of $\sim \mO(10^{-24}) - \mO(10^{-21})$ (Type-I) or $\sim \mO(10^{-24}) - \mO(10^{-22})$ (Type-II) have been ruled out in the collapsing domain wall scenario.
Another order of magnitude from the future improvements to the eEDM will correspondingly constrain the tiny ECPV parameter in the extended ranges of $\sim \mO(10^{-24}) - \mO(10^{-20})$ (Type-I) or $\sim \mO(10^{-24}) - \mO(10^{-21})$ (Type-II).


\section{Conclusion and Discussion}
\label{section:conclusion}

In this work, we focus on the vacuum with the SCPV in the 2HDM, which can lead to a domain wall structure.
Through our numerical studies, such domain walls typically lead to incredibly large tensions of $\mO(10^6)\,\GeV^3$, which is well above the Zel'dovich-Kobzarev-Okun bound.
Therefore, the complex parameters in the 2HDM potential are necessary in playing the role as the biased terms to collapse these domain walls.
With reasonable sizes of the biased terms, such domain walls could have been formed in the early Universe.
In order not to spoil the BBN constraints, we find the direct constraints to the sizes of the ECPV terms, for which we choose to be ${\rm Im} \lambda_5$, to be $\gtrsim \mO( 10^{-14}- 10^{-12})$ ($\gtrsim \mO( 10^{-24}- 10^{-22}$) with the CPV mixing angles of $\alpha_c\sim \mO(10^{-10})$ ($\alpha_c\sim \mO(1)$).
Although the related domain wall collapse does not lead to sufficiently strong signals for the future GW probes at the SKA, we find that the eEDM measurements will play a role to probe the deep echos from this process.
The future projection of the eEDM measurements from the ACME-III can set upper limits to the CPV mixing angle of $\alpha_c \lesssim \mO(10^{-2}) - \mO(10^{-4})$, which depends on the types of the Yuakwa couplings and mass spliting of two neutral heavy Higgs bosons.
For the first time, we find the current and future eEDM measurements have excluded or can be used to probe the very tiny regions of the ECPV parameter of ${\rm Im} \lambda_5 \sim \mO( 10^{-24} ) - \mO(10^{-20}) $ (Type-I) or ${\rm Im} \lambda_5 \sim \mO( 10^{-24} ) - \mO(10^{-21}) $ (Type-II).
In other words, we find that the eEDM measurements can look deep into the possible domain wall collapses in the early Universe.
This is different from the discussions in the context of SUSY where the improved eEDM measurements were thought to set lower limit to the SUSY-breaking scale.
In the context of the domain wall collapsing triggered by the complex parameters in the 2HDM potential, the future improvements of the eEDM measurements are found to further set up upper limits to the CPV mixing angles and the sizes of the ECPV term.

Some future efforts can be envisioned from this work.

\begin{enumerate}

\item The 2HDM, along with other new physics models, are known to produce other topological defects besides of the domain wall solutions from the CP symmetry, such as vortices and monopoles~\cite{La:1993je,Earnshaw:1993yu,Bimonte:1994qh,Battye:2011jj,brawn_SymmetriesTopologicalDefects_2011,Eto:2019hhf,Eto:2020hjb}.
The solutions to these structure can lead to GW signals as well.
Therefore, it will be useful to probe the complete spectrum of the GWs for given new physics model.

\item We did not consider the electroweak phase transition and the possibility of achieving the BAU within this frame.
It was known that various topological defects, such as monopoles, may play a role of producing the baryon number violation process~\footnote{See early Refs.~\cite{Rubakov:1981rg,Callan:1982au,Callan:1982ac} for the Rubakov-Callan effects.}.
There have been some recent progresses~\cite{Cline:2020jre,Zhou:2020xqi,Xie:2020bkl} along this direction.
It is therefore to ask if a successful BAU can be achieved when the new physics models have non-trivial topological solutions.

\item From the perspective of the EDM measurements, it will be also useful to perform the estimations to the atomic EDMs and find the future experimental search projections as well.

\end{enumerate}


\section*{ACKNOWLEDGMENTS}

We would like to thank Yuan Sun, Tian Xia, Ke-Pan Xie and Yue Zhang for very useful discussions and communication. The work of NC is partially supported by the National Natural Science Foundation of China (under Grant No. 11575176).
TL is supported by the National Natural Science Foundation of China (Grant No. 11975129) and ``the Fundamental Research Funds for the Central Universities'', Nankai University (Grant No. 63196013).
The work of NC and TL is also supported in part by the National Natural Science Foundation of China (Grant No. 12035008).
YW is supported by the Natural Sciences and Engineering Research Council of Canada (NSERC).

\newpage

\appendix

\section{The mass spectrum and self couplings with the SCPV and the ECPV}
\label{section:2HDMspectrum}

In this appendix, we provide explicit relations between the 2HDM parameters in the generical basis and the physical basis. 
Many details of deriving the relative CPV phase $\theta$ in the SCPV and the ECPV scenarios are also presented.

\subsection{The mass spectrum and self couplings with the SCPV}
\label{section:2HDMSCPV}

For the SCPV scenario, we have the charged Higgs boson mass squared matrix of
\begin{align}
&\mM_\pm^2/v^2 =- \hf ( \lambda_4 - {\rm Re} \lambda_5 ) \left( \ba{cc}
 s_\beta^2 &     -s_\beta c_\beta e^{i \theta }  \\
 - s_\beta c_\beta e^{-i \theta } & c_\beta^2    \\  \ea \right)\,,
\end{align}
and the corresponding eigenvalues are
\beqn\label{eqs:chargedH_SCPV}
m_\pm^2&=& - \hf ( \lambda_4 - {\rm Re} \lambda_5 ) v^2  \,.
\eeqn
The gauge eigenstates of $(H_1^\pm \,, H_2^\pm )$ are transformed into mass eigenstates of charged Nambu-Goldstone bosons $G^\pm$ and Higgs bosons $H^\pm$ by
\beqn\label{eq:chargedH_diagonal}
\left(  \ba{c}  H_1^\pm \\ H_2^\pm  \ea  \right)&=&\left(  \ba{cc}
c_\beta & - s_\beta  \\
s_\beta  & c_\beta  \\  \ea  \right) \cdot    \left(  \ba{c}  G^\pm \\ H^\pm   \ea  \right)\,.
\eeqn
Similarly, two pseudoscalars of $(A_1\,, A_2)$ are transformed into neutral Nambu-Goldstone bosons $G^0$ and a pseudoscalar $A^0$ by
\beqn\label{eq:neutralA_diagonal}
\left(  \ba{c}  A_1   \\  A_2  \ea  \right)&=&\left(  \ba{cc}
c_\beta & - s_\beta  \\
s_\beta  & c_\beta  \\  \ea  \right) \cdot    \left(  \ba{c}  G^0 \\ A^0   \ea  \right)\,.
\eeqn

In the basis of $( H_1\,, H_2\,, A^0)$, the $3\times 3$ mass squared matrix for the neutral states are diagonalized by~\cite{Bimonte:1994qh,WahabElKaffas:2007xd,Inoue:2014nva}
\beqs
\label{eq:rotation_R}
\beqn
\left(  \ba{c}  h_1  \\  h_2 \\ h_3   \ea  \right)&=& \mR \cdot \left(\ba{c}  H_1 \\  H_2 \\ A^0  \ea  \right)
\,,\\
\mM_0^2&=& \mR^T \cdot   \left( \ba{ccc}
m_1^2 &   0   & 0   \\
0  &  m_2^2  & 0  \\
0 &  0 &   m_3^2 \\    \ea  \right) \cdot \mR \,,\\
\mR&=& \mR_{23} (\alpha_c) \cdot \mR_{13}(\alpha_b ) \cdot \mR_{12} (\alpha+ \frac{\pi}{2} ) \non
&=& \left(\begin{array}{ccc}
        1 & 0 & 0 \\
        0 & c_{\alpha_c} & s_{\alpha_c} \\
        0 & -s_{\alpha_c} & c_{\alpha_c}
        \end{array}\right)\cdot
        \left(\begin{array}{ccc}
        c_{\alpha_b} & 0 & s_{\alpha_b} \\
        0 & 1 & 0 \\
        -s_{\alpha_b} & 0 & c_{\alpha_b}
        \end{array}\right)\cdot
        \left(\begin{array}{ccc}
        -s_\alpha & c_\alpha & 0 \\
        -c_\alpha & -s_\alpha & 0 \\
        0 & 0 & 1
        \end{array}\right) \non
&=&\left( \ba{ccc}
- s_\alpha c_{\alpha_b}  &  c_\alpha c_{\alpha_b}  &  s_{\alpha_b}  \\
s_\alpha s_{\alpha_b } s_{\alpha_c } - c_\alpha c_{\alpha_c}  &  -s_\alpha c_{\alpha_c} - c_\alpha s_{\alpha_b} s_{\alpha_c}   &  c_{\alpha_b}  s_{\alpha_c} \\   
s_\alpha s_{\alpha_b} c_{\alpha_c } + c_\alpha s_{\alpha_c}  &  s_\alpha s_{\alpha_c} - c_\alpha s_{\alpha_b} c_{\alpha_c}   &  c_{\alpha_b} c_{\alpha_c}  \\  \ea \right)\,.
\eeqn
\eeqs
Each term in the mass squared matrix of $\mM_0^2$ is listed below
\begin{subequations}\label{eqs:MN2_SCPV}
\begin{align}
\mM_0^2/v^2 &= \left( \ba{ccc}
   \hat \mu_{11}^2 &   \hat \mu_{12}^2  & \hat \mu_{1A}^2     \\
  \hat \mu_{12}^2  &   \hat \mu_{22}^2 & \hat \mu_{2A}^2 \\
\hat \mu_{1A}^2  & \hat \mu_{2A}^2  &  \hat \mu_{AA}^2  \\  \ea   \right) \,,\label{eq:MN2_matrix}\\
\hat \mu_{11}^2 &= \lambda_1 c_\beta^2 + {\rm Re} \lambda_5 s_\beta^2 c_{\theta }^2 \,,\\
\hat \mu_{22}^2 &= \lambda_2 s_\beta^2  + {\rm Re} \lambda_5 c_\beta^2 c_{\theta}^2 \,,\\
\hat \mu_{12}^2 &= ( \lambda_3 + \lambda_4 - {\rm Re} \lambda_5 s_{\theta }^2   ) s_\beta c_\beta   \,,\\
\hat \mu_{1A}^2 &= - \hf {\rm Re} \lambda_5 s_{2\theta }  s_\beta \,,\\
\hat \mu_{2A}^2 &= - \hf {\rm Re} \lambda_5   s_{2\theta }  c_{\beta }     \,,\\
\hat \mu_{AA}^2 &= {\rm Re} \lambda_5   s_{ \theta }^2 \,.
\end{align}
\end{subequations}
In turn, the quartic scalar self couplings are expressed in terms of the Higgs boson masses and mixings as
\beqs\label{eqs:SCPV_lambda}
\beqn
\lambda_1&=& \frac{1}{ c_\beta^2 } \sum_{i=1}^3  \frac{ m_i^2}{v^2} ( \mR_{i1}^2 - \frac{ s_\beta^2 }{ t_{\theta}^2 } \mR_{i3}^2 )\,, \\
\lambda_2&=& \frac{1}{ s_\beta^2 } \sum_{i=1}^3 \frac{ m_i^2}{v^2} ( \mR_{i2  }^2 - \frac{ c_\beta^2 }{ t_{\theta}^2 } \mR_{i3}^2 )\,, \\
\lambda_{3} &=&  \sum_{i=1}^3  \frac{ m_i^2}{v^2} ( \frac{1}{ s_\beta c_\beta } \mR_{i1} \mR_{i2} - \frac{1}{ t_\theta^2 }  \mR_{i3}^2   )  + \frac{2  m_\pm^2}{v^2} \,,\\
\lambda_4 &=&  \frac{1}{ s_\theta^2} \sum_i  \frac{ m_i^2}{v^2} \mR_{i3}^2  - \frac{ 2  m_\pm^2 }{v^2}\,,\\
{\rm Re} \lambda_5 &=& \frac{1}{ s_\theta^2} \sum_i  \frac{ m_i^2}{v^2} \mR_{i3}^2  \,,
\eeqn
\eeqs
in the SCPV scenario.

There is a well-known constraint from the $(\mM_0^2)_{13}$ and $(\mM_0^2)_{13}$ terms in Eq.~\eqref{eqs:MN2_SCPV}:
\beqn
&& (\mM_0^2)_{13} = t_{ \beta}\, (\mM_0^2)_{23}\,.
\eeqn
This leads to one additional constraint between mixing angles and mass eigenvalues as follows~\cite{Khater:2003wq}
\beqn\label{eq:mass_constraint}
&&( m_1^2 - m_2^2 s_{\alpha_c}^2  - m_3^2 c_{\alpha_c}^2 ) s_{\alpha_b} (  t_\alpha + t_\beta )= ( m_2^2 - m_3^2 ) (t_\alpha t_{\beta} - 1) s_{\alpha_c} c_{\alpha_c}\non
&\Rightarrow& s_{\alpha_b} = \frac{ ( m_2^2 - m_3^2 ) (t_\alpha t_{\beta} - 1) s_{\alpha_c} c_{\alpha_c} }{ ( m_1^2 - m_2^2 s_{\alpha_c}^2  - m_3^2 c_{\alpha_c}^2 )  (  t_\alpha + t_\beta ) }  \,.
\eeqn
In practice, we use three mixing angles of $(\alpha\,,\alpha_c\,,\beta)$ as inputs.
With the special limit of $\beta-\alpha=\pi/2$, we have
\beqn\label{eq:mass_constraint_spec}
&&( m_1^2 - m_2^2 s_{\alpha_c}^2  - m_3^2 c_{\alpha_c}^2 ) s_{\alpha_b} = ( m_2^2 - m_3^2 )  s_{\alpha_c} c_{\alpha_c} t_{ 2\beta} \non
&\Rightarrow & s_{\alpha_b } =  \frac{ ( m_2^2 - m_3^2 ) s_{\alpha_c} c_{\alpha_c} t_{2\beta } }{ m_1^2 - m_2^2 s_{\alpha_c}^2 - m_3^2 c_{\alpha_c}^2 } \,,
\eeqn
and there is a singularity at $t_\beta=1$ under this limit.
In the small $\alpha_c$ limit, we can further approximate this relation as
\beqn\label{eq:alphab_Approx}
\alpha_b&\approx&  \frac{ m_3^2 - m_2^2 }{ m_3^2 - m_1^2 } t_{2\beta} \alpha_c \,.
\eeqn

In the physical basis, the SCPV phase of $\theta$ is obtained from the relation of
\beqn\label{eq:theta_SCPV}
&& \frac{ \hat \mu_{AA}^2 }{ \hat \mu_{1A}^2 } = - \frac{ s_\beta }{ t_\theta }  \Rightarrow t_\theta= -s_\beta \frac{ \sum_i  m_i^2 \mR_{i3}^2 }{   \sum_i  m_i^2 \mR_{i1} \mR_{i3}  } \,.
\eeqn
One may express $\hat \mu_{1A}^2$ and $ \hat \mu_{AA}^2$ in terms of masses and mixing angles explicitly as follows
\beqn
\hat \mu_{1A}^2 v^2 &=& ( m_2^2 s_{\alpha_c}^2 + m_3^2 c_{\alpha_c}^2 - m_1^2 ) s_\alpha s_{\alpha_b } c_{\alpha_b } + ( m_3^2 - m_2^2 ) c_\alpha c_{\alpha_b} s_{\alpha_c } c_{\alpha_c}\;, \non
\hat \mu_{AA}^2 v^2 &=&  m_1^2 s_{\alpha_b}^2 +( m_2^2 s_{\alpha_c}^2  + m_3^2 c_{\alpha_c}^2 ) c_{\alpha_b}^2  \,.
\eeqn
By combining with the dependences of $|\alpha_b|$ on $\alpha_c$ as depicted in Fig.~\ref{fig:alphab}, the $\hat \mu_{1A}^2$ and $ \hat \mu_{AA}^2$ become
\beqn
&& \hat \mu_{1A}^2  \sim  - \frac{  m_3^2  - m_2^2 }{v^2} \frac{ s_\beta \alpha_c  }{c_{2\beta} } \,,\qquad  \hat \mu_{AA}^2  \sim    \frac{m_3^2 }{v^2}  \,.
\eeqn
under the $\beta-\alpha=\pi/2$ and small $\alpha_c$ limit.
Here, we have also made used of Eq.~\eqref{eq:alphab_Approx}.
Clearly, one can envision that $\hat \mu_{1A}^2 \ll \hat \mu_{AA}^2$.
Accordingly, the SCPV phase becomes
\beqn
t_\theta &\approx& \frac{ m_3^2 c_{2\beta} }{ ( m_3^2  - m_2^2 ) \alpha_c } \,.
\eeqn
Obviously, the solutions of the SCPV are approaching to $\pm \frac{\pi}{2} $ when the CPV mixing angle $\alpha_c$ is suppressed and/or two heavy neutral Higgs bosons are very mass degenerate.

\begin{table}[htb]
\begin{center}
\begin{tabular}{c|c}
\hline
\hline
 Generical basis  &  Physical basis  \\
 \hline
$\lambda_{1\,,2\,,3\,,4}\,, {\rm Re} \lambda_5$  &  $m_{1\,,2\,,3}$\,, $m_\pm$\,, $v$  \\
 $m_{11}^2\,, m_{22}^2\,, {\rm Re} m_{12}^2$  &  $\alpha\,,\alpha_b\,, \alpha_c\,, \beta\,,\theta$ \\
\hline\hline
\end{tabular}
\caption{
The parameter inputs in both the generical basis and the physical basis for the SCPV scenario.}
\label{tab:SCPV_parameters}
\end{center}
\end{table}

We summarize the parameter inputs for the SCPV scenario in both the physical basis and the generical basis in Table.~\ref{tab:SCPV_parameters}.
The CPV mixing angle $\alpha_b$ and the SCPV phase $\theta$ can be obtained by Eq.~\eqref{eq:mass_constraint} and Eq.~\eqref{eq:theta_SCPV}, respectively.
Therefore, one has eight independent parameters in both basis.

\subsection{The mass spectrum and self couplings with the ECPV}
\label{section:2HDMECPV}

The most general 2HDM potential with complex parameters of $m_{12}^2$ and $\lambda_5$ violates the CP symmetry explicitly.
The corresponding minimization conditions were previously given in Eqs.~\eqref{eqs:ECPV_min}.

For the mass spectrum, the same conventions of the mass squared matrix and mixing angles are adopted as in the SCPV scenario.
The charged Higgs boson masses are expressed as
\beqn
m_\pm^2 &=& - \frac{1}{2}\left( \lambda_4 - {\rm Re}\lambda_5 - {\rm Im} \lambda_5/t_\theta  \right) v^2 - \frac{{\rm Im} m_{12}^2}{s_\beta c_\beta s_\theta} \,.
\eeqn
Schematically, the neutral mass squared matrix is the same as in Eq.~\eqref{eq:MN2_matrix}, with each element listed as below
\beqs\label{eqs:MN2_ECPV}
\beqn
 \hat \mu_{11}^2 &=& \lambda_1 c_\beta^2 + {\rm Re} \lambda_5 c_\theta^2s_\beta^2 + {\rm Im} \lambda_5 \frac{  c_{2\theta}s_\beta^2}{2t_\theta} -  \frac{ {\rm Im} m_{12}^2\, t_\beta }{v^2 s_\theta } \,,\\
 \hat \mu_{22}^2 &=& \lambda_2 s_\beta^2 + {\rm Re}  \lambda_5  c_\theta^2 c_\beta^2 + {\rm Im} \lambda_5 \frac{  c_{2\theta}c_\beta^2}{2 t_\theta} -  \frac{ {\rm Im} m_{12}^2}{v^2 t_\beta s_\theta} \,,\\
\hat \mu_{12}^2  &=& (\lambda_3 + \lambda_4 - {\rm Re}  \lambda_5 s_\theta^2 - {\rm Im} \lambda_5 \frac{1+2s_\theta^2}{2t_\theta}  ) s_\beta c_\beta + \frac{ {\rm Im} m_{12}^2}{v^2 s_\theta}\,,\\
\hat \mu_{1A}^2 &=& -\frac{1}{2} \left({\rm Re}  \lambda_5s_{2\theta} + {\rm Im} \lambda_5 c_{2\theta}\right) s_\beta \,, \\
\hat \mu_{2A}^2 &=& -\frac{1}{2} \left({\rm Re}  \lambda_5s_{2\theta} + {\rm Im} \lambda_5c_{2\theta}\right) c_\beta \,,\\
\hat \mu_{AA}^2  &=& {\rm Re}  \lambda_5 s_\theta^2 + {\rm Im} \lambda_5 \frac{1+2s_\theta^2}{2t_\theta} - \frac{ {\rm Im} m_{12}^2}{ v^2 s_\beta c_\beta s_\theta } \,.
\eeqn
\eeqs
From the expressions of $\hat \mu_{1A}^2$ and $\hat \mu_{AA}^2$, one finds the relation between the imaginary components of ${\rm Im}  m_{12}^2$ and ${\rm Im} \lambda_5$ as
\beqn\label{eq:thetaSol_ECPV}
&&  \hat \mu_{AA}^2 s_\theta c_\theta + \frac{ \hat \mu_{1A}^2 }{ s_\beta} s_\theta^2 =  \hf {\rm Im}  \lambda_5  - \frac{ {\rm Im}  m_{12}^2}{ v^2 s_\beta c_\beta } c_\theta  \,,
\eeqn
which will be used for solving the relative CPV phase of $\theta$ in general when replacing $\hat\mu_{AA}^2$ and $\hat\mu_{1A}^2$ by physical inputs according to Eq.~\eqref{eq:rotation_R}.

In the ECPV scenario, the phase transformation of $\Phi_2$ can remove one of the three phases.
The third minimization condition in Eq.~\eqref{eq:ECPV_min03} will help to relate the two remaining phases.
Thus, we have only one free CPV phase again.
There are several situations in solving the CPV phase of $\theta$ from Eq.~\eqref{eq:thetaSol_ECPV} for the ECPV scenario:
\begin{enumerate}

    \item ${\rm Im} m_{12}^2 = 0$, the above equation is a quadratic equation of $t_\theta$.
    This is equivalent to rephase $\Phi_2$ as $\Phi_2 \to e^{-i \delta_1 } \Phi_2$.

    \item ${\rm Im} \lambda_5 = 0$, the above equation is a quartic equation of $s_\theta$.
    This is equivalent to rephase $\Phi_2$ as $\Phi_2 \to e^{-i \delta_2/2 } \Phi_2$.

    \item If one keeps both ${\rm Im} m_{12}^2$ and ${\rm Im} \lambda_5$ non-zero, the above equation is a quartic equation of $t_{\theta/2}$.

\end{enumerate}
To simplify our discussions, we take the special case of ${\rm Im} m_{12}^2 = 0$, where $t_\theta$ satisfies
\beqn
&& \left(\frac{2 \hat \mu_{1A}^2 }{s_\beta}- {\rm Im} \lambda_5 \right)t_\theta^2 +  2 \hat \mu_{AA}^2 t_\theta -  {\rm Im} \lambda_5 = 0\,.
\eeqn
The exact solution for $t_\theta$ is expressed as
\beqn\label{eq:ECPV_theta_ExactSol}
t_\theta &=& \frac{1}{ 2\hat \mu_{1A}^2 / s_\beta - {\rm Im} \lambda_5 } \Big[ \mp \sqrt{ ( \hat \mu_{AA}^2)^2 + {\rm Im} \lambda_5 ( 2 \hat \mu_{1A}^2/s_\beta  - {\rm Im} \lambda_5 )  }  - \hat \mu_{AA}^2 \Big] \,.
\eeqn
Given that we have $\hat \mu_{1A}^2 \ll \hat \mu_{AA}^2$ in the small $\alpha_c$ limit, the $\mp$ signs in the exact solutions in Eq.~\eqref{eq:ECPV_theta_ExactSol} correspond to the limits of $\theta \to \pm \pi/2$ and $\theta\to 0$, respectively.
To restore the exact solution of $t_\theta$ in the SCPV scenario, one should take the $-$ sign in Eq.~\eqref{eq:ECPV_theta_ExactSol}.
It turns out that the relative CPV phase $\theta$ will enter into the biased term to make the domain wall collapse in the form of $s_{2\theta}$.
In both limit of $\theta \to \pm \pi/2$, we have $s_{2 \theta}\to 0$.

Besides of the CPV mixing angle of $\alpha_c$, the ECPV parameter of ${\rm Im} \lambda_5$ also plays a role.
In the limit of $|{\rm Im} \lambda_5| \ll \hat \mu_{1A}^2/s_\beta $, we have
\beqn\label{eq:ECPV_thetaApprox01}
t_\theta &\approx& -s_\beta \frac{\hat \mu_{AA}^2 }{\hat \mu_{1A}^2 } - \frac{{\rm Im} \lambda_5 s_\beta}{2 \hat \mu_{1A}^2 }\left( \frac{s_\beta  \hat \mu_{AA}^2 }{ \hat \mu_{1A}^2 } + \frac{  \hat \mu_{1A}^2 }{s_\beta   \hat \mu_{AA}^2 }\right) + \mathcal{O}\left(\left( {\rm Im} \lambda_5\right)^2\right) \,,
\eeqn
with the leading contributions the same as the expression in the SCPV scenario Eq.~\eqref{eq:theta_SCPV}.
Correspondingly, the relative CPV phase of $\theta$ is expected to depend on CPV mixing angle $\alpha_c$ significantly.
On the other hand, in the limit of $ \hat \mu_{AA}^2 \gg |{\rm Im} \lambda_5| \gg \hat \mu_{1A}^2/s_\beta $, we have the approximation to $t_\theta$ as
\beqn\label{eq:ECPV_thetaApprox02}
t_\theta &\approx& \frac{2 \hat \mu_{AA}^2}{{\rm Im} \lambda_5} - \frac{ {\rm Im} \lambda_5 }{ 2 \hat \mu_{AA}^2} +  \mathcal{O}\left(\left({\rm Im} \lambda_5\right)^2\right)\,.
\eeqn
As we have shown above that $\hat \mu_{AA}^2 \to m_3^2/v^2$ in the small $\alpha_c$ limit, the value of $\theta$ will be independent of the CPV mixing angle $\alpha_c$ in such a limit.

With the solution of $\theta$ obtained in Eq.~\eqref{eq:ECPV_theta_ExactSol}, we can solve for the quartic scalar self couplings in terms of the Higgs boson masses and mixings as below
\beqs\label{eqs:ECPV_lambda}
\beqn
\lambda_1 &=&  \sum_{i=1}^3\frac{m_i^2}{v^2  c_\beta^2 }\left( \mR_{i1}^2-\frac{s_\beta^2}{t_\theta^2} \mR_{i3}^2\right) + \frac{ {\rm Im} \lambda_5\,t_\beta^2}{2 t_\theta s_\theta^2} \,,\\
\lambda_2 &=& \sum_{i=1}^3\frac{m_i^2}{v^2  s_\beta^2}\left( \mR_{i2}^2-\frac{c_\beta^2}{t_\theta^2} \mR_{i3}^2\right) + \frac{ {\rm Im} \lambda_5 }{2 t_\beta^2t_\theta s_\theta^2}\,,\\
\lambda_3 &=&\sum_{i=1}^3\frac{m_i^2}{ v^2  s_\beta c_\beta }\left( \mR_{i1} \mR_{i2} - \frac{s_\beta c_\beta}{t_\theta^2} \mR_{i3}^2\right)  + \frac{2m_\pm^2}{v^2} + \frac{ {\rm Im} \lambda_5 }{2 t_\theta s_\theta^2} \,,\\
\lambda_4 &=& \sum_{i=1}^3\frac{m_i^2 \mR_{i3}^2}{v^2 s_\theta^2 } -\frac{2m_\pm^2}{v^2}  -\frac{{\rm Im} \lambda_5}{2 t_\theta s_\theta^2}  \,,\\
{\rm Re} \lambda_5  &=& \sum_{i=1}^3\frac{m_i^2 \mR_{i3}^2}{v^2 s_\theta^2 } -{\rm Im }\lambda_5\frac{1+2s_\theta^2}{2 t_\theta s_\theta^2}  \,.
\eeqn
\eeqs
with the choice of ${\rm Im} m_{12}^2=0$.
All quartic scalar self couplings can easily return to those in the SCPV scenario as listed in Eqs.~\eqref{eqs:SCPV_lambda}, by taking ${\rm Im} \lambda_5 = 0$. 
The Higgs potential with the ECPV scenario can be extended from Eq.~\eqref{eqs:SCPV1_potential} as below
\beqs\label{eqs:CPV_potentials}
\beqn
V&=& V_{\rm CPC} + V_{\rm CPV}  \,,\\
V_{\rm CPV}&=& V_{\rm SCPV} + V_{\rm ECPV} \,,\\
V_{\rm SCPV}&=& \Big( - \frac{ {\rm Re}  m_{12}^2  }{ v^2\, s_\beta c_\beta } c_\theta + \frac{  1 }{4} {\rm Re} \lambda_5 c_{2\theta} \Big) v^4 s_\beta^2 c_\beta^2 \,,\\
V_{\rm ECPV}&=&  -\frac{ 1 }{4} {\rm Im}\lambda_5 s_{2 \theta }    v^4 s_\beta^2 c_\beta^2 \,.
\eeqn
\eeqs
%

\begin{table}[htb]
\begin{center}
\begin{tabular}{c|c}
\hline
\hline
  Generical basis &  Physical basis  \\
 \hline
$\lambda_{1\,,2\,,3\,,4}\,, {\rm Re} \lambda_5 \,, {\rm Im} \lambda_5 $  &  $m_{1\,,2\,,3}$\,, $m_\pm$\,, $v$  \\
 $m_{11}^2\,, m_{22}^2\,, {\rm Re} m_{12}^2\,, {\rm Im} m_{12}^2$  &  $\alpha\,,\alpha_b\,, \alpha_c\,, \beta\,,\theta\,,{\rm Im}\lambda_5\,, {\rm Im} m_{12}^2 $ \\
\hline\hline
\end{tabular}
\caption{
The parameter inputs in both the generical basis and the physical basis for the ECPV scenario.
}
\label{tab:ECPV_parameters}
\end{center}
\end{table}

We summarize the parameter inputs for the ECPV scenario in both the physical basis and the generical basis in Table.~\ref{tab:ECPV_parameters}.
The CPV mixing angle $\alpha_b$ is obtained from Eq.~\eqref{eq:mass_constraint} from other masses and mixing angles.
The relative CP phase can be obtained from Eq.~\eqref{eq:ECPV_theta_ExactSol}.
As we have argued previously, one can always remove one of two ECPV parameters by rephasing the second doublet $\Phi_2$.
Without loss of generality, we choose to take ${\rm Im} m_{12}^2=0$.
Hence, one has nine independent input parameters in both basis.
We shall describe the procedures of converting the input parameters in the physical basis to the generic basis.
\begin{enumerate}

\item In the physical basis, all necessary input parameters are masses of three neutral Higgs bosons $m_{1\,,2\,,3}$, the mass of charged Higgs bosons $m_\pm$, the Higgs VEV of $v= (\sqrt{2} G_F)^{-1/2}\simeq 246\,\GeV$, and three mixing angles of $(\alpha\,, \alpha_c\,,\beta)$, as well as the ECPV parameter ${\rm Im}\lambda_5$.
The state of $h_1$ will be regarded as the SM-like Higgs boson, with $m_1=125\,\GeV$.
To simplify the discussion, we shall take the special limit of $\alpha= \beta - \frac{\pi }{2}$.

\item The other CPV mixing angle $\alpha_b$ is related to three input mixing angles of $(\alpha \,, \alpha_c\,, \beta)$ from Eq.~\eqref{eq:mass_constraint}, which is reduced to Eq.~\eqref{eq:mass_constraint_spec} in the special limit of $\alpha= \beta - \frac{\pi }{2}$.

\item The relative CPV phase of $\theta$ is derived from the exact solution in Eq.~\eqref{eq:ECPV_theta_ExactSol} with all above parameter plus the ECPV parameter of ${\rm Im} \lambda_5$.

\item With all Higgs masses of $m_{1\,,2\,,3\,,\pm}$, mixing angles of $(\alpha\,,\alpha_b\,,\alpha_c\,,\beta)$, the relative CPV phase of $\theta$, and the ECPV parameter of ${\rm Im} \lambda_5$, we obtain the quartic scalar self couplings from Eqs.~\eqref{eqs:ECPV_lambda}, and mass squared parameters of $(m_{11}^2 \,, m_{22}^2\,, {\rm Re} m_{12}^2 )$ from the minimization conditions in Eqs.~\eqref{eqs:ECPV_min}.
These parameters in the generical basis will be used for solving the domain wall profiles and determine the sizes of the biased terms for the domain wall collapse.

\end{enumerate}

\subsection{The Yukawa couplings in the CPV 2HDM}
\label{section:2HDM_Yukawa}

\begin{table}[htb]
\begin{center}
\begin{tabular}{c|c|c}
\hline
\hline
 & Type-I & Type-II   \\\hline
 $c_{u\,,1}$  &  $ \mR_{12}/s_\beta \to c_{\alpha_b}$  & $ \mR_{12}/s_\beta \to c_{\alpha_b}$  \\
 $c_{d\,,1}=c_{\ell\,,1}$  & $ \mR_{12}/s_\beta \to  c_{\alpha_b}$     &  $ \mR_{12}/s_\beta \to c_{\alpha_b}$   \\
 $\tilde c_{u\,,1}$  & $-\mR_{13}/t_\beta \to -s_{\alpha_b}/t_\beta$   &  $ -\mR_{13}/t_\beta \to -s_{\alpha_b}/t_\beta$    \\
 $\tilde c_{d\,,1}=\tilde c_{\ell\,,1}$  &  $\mR_{13}/t_\beta \to s_{\alpha_b}/t_\beta$   &  $  -\mR_{13} t_\beta \to -s_{\alpha_b}\,t_\beta$   \\
 $a_1$  &  $ \mR_{12} s_\beta + \mR_{11} c_\beta \to  c_{\alpha_b}$  &   $ \mR_{12} s_\beta + \mR_{11} c_\beta \to c_{\alpha_b}$  \\
 \hline
 $c_{u\,,2}$  & $ \mR_{22}/s_\beta \to c_{\alpha_c}/t_\beta - s_{\alpha_b} s_{\alpha_c}$   &  $ \mR_{22}/s_\beta \to c_{\alpha_c}/t_\beta - s_{\alpha_b} s_{\alpha_c}$   \\
 $c_{d\,,2}=c_{\ell\,,2}$  &  $\mR_{22}/s_\beta \to c_{\alpha_c}/t_\beta - s_{\alpha_b} s_{\alpha_c}$   & $ \mR_{21}/c_\beta \to - s_{\alpha_b} s_{\alpha_c} - c_{\alpha_c}\,t_\beta  $    \\
 $\tilde c_{u\,,2}$  &  $ -\mR_{23}/t_\beta \to -c_{\alpha_b}s_{\alpha_c}/t_\beta$   &   $-\mR_{23}/t_\beta \to -c_{\alpha_b}s_{\alpha_c}/t_\beta$  \\
 $\tilde c_{d\,,2}=\tilde c_{\ell\,,2}$  &  $ \mR_{23}/t_\beta \to c_{\alpha_b}s_{\alpha_c}/t_\beta$   &  $-\mR_{23} t_\beta \to -c_{\alpha_b}\,s_{\alpha_c}\,t_\beta$   \\
 $a_2$  &  $\mR_{22} s_\beta + \mR_{21} c_\beta \to -s_{\alpha_b} s_{\alpha_c}  $  & $\mR_{22} s_\beta + \mR_{21} c_\beta \to -  s_{\alpha_b} s_{\alpha_c} $  \\
 \hline
 $c_{u\,,3}$  & $ \mR_{32}/s_\beta \to - s_{\alpha_c}/t_\beta  -  s_{\alpha_b} c_{\alpha_c} $   &  $\mR_{32}/s_\beta \to - s_{\alpha_c}/t_\beta  -  s_{\alpha_b} c_{\alpha_c} $   \\
 $c_{d\,,3}=c_{\ell\,,3}$  &  $ \mR_{31}/s_\beta\to  - s_{\alpha_c}/t_\beta  -  s_{\alpha_b} c_{\alpha_c}$   &  $ \mR_{31}/c_\beta\to - s_{\alpha_b} c_{\alpha_c}+ s_{\alpha_c}\, t_\beta $   \\
 $\tilde c_{u\,,3}$  &  $ -\mR_{33}/t_\beta \to -c_{\alpha_b}c_{\alpha_c}/t_\beta$  &  $ -\mR_{33}/t_\beta \to -c_{\alpha_b}c_{\alpha_c}/t_\beta$   \\
 $\tilde c_{d\,,3}=\tilde c_{\ell\,,3}$  &  $ \mR_{33}/t_\beta \to c_{\alpha_b}c_{\alpha_c}/t_\beta$   &  $ -\mR_{33}t_\beta \to -c_{\alpha_b}c_{\alpha_c}\, t_\beta$   \\
 $a_3$  &   $ \mR_{32} s_\beta + \mR_{31} c_\beta \to - s_{\alpha_b} c_{\alpha_c} $ &   $ \mR_{32} s_\beta + \mR_{31} c_\beta \to  -  s_{\alpha_b} c_{\alpha_c} $   \\
 \hline\hline
\end{tabular}
\caption{
The SM fermion and gauge boson couplings to Higgs boson mass eigenstates of $h_{1\,,2\,,3}$, and their expressions in the limit of $\beta-\alpha = \pi/2$.
 }\label{tab:Hcouplings}
\end{center}
\end{table}

We focus on the CPV 2HDM where the Yukawa sector has a $\mathbb{Z}_2$ symmetry and $\Phi_1$ and $\Phi_2$ each only gives mass to up-type quarks or down-type quarks and charged leptons.
This is sufficient to suppress tree-level flavor changing processes mediated by the neutral Higgs bosons.
The Yukawa couplings for the Type-I and Type-II 2HDM read (and suppressing the CKM mixing),
\beqn\label{eq:2HDM_Yuk}
\mathcal{L}= \left\{\begin{array}{ll}
-( { \frac{c_\alpha}{s_\beta}}{\frac{m_u}{v}} )\overline Q_L \tilde \Phi_2 u_R - (  {  \frac{c_\alpha}{s_\beta}}{\frac{m_d}{v}} ) \overline Q_L \Phi_2 d_R + {\rm h.c.} & \hspace{1cm} {\rm Type-I}\vspace{0.2cm} \\
- (  { \frac{c_\alpha}{s_\beta}}{\frac{m_u}{v}}) \overline Q_L \tilde \Phi_2 u_R
+( { \frac{s_\alpha}{c_\beta}}{\frac{m_d}{v}} )\overline Q_L \Phi_1 d_R
+ {\rm h.c.} & \hspace{1cm} {\rm Type-II} \, ,
\end{array} \right.
\eeqn
where $Q_L^T=(u_L,d_L)$ and $\tilde \Phi_2 \equiv i \sigma_2 \Phi_2^*$.
For both cases, the charged lepton Yukawa coupling has the same form as that of the down-type quarks.
Therefore, we can express the couplings between neutral Higgs bosons and the fermions and gauge bosons in the mass eigenbasis
\beqn\label{coup_f}
\mL&=& \sum_{i=1}^3 \left[-m_f\left( c_{f,i} \bar f f+ \tilde c_{f,i} \bar f i\gamma_5 f  \right) + a_i \left( 2  m_W^2 W_\mu W^\mu +  m_Z^2 Z_\mu Z^\mu \right)  \right] \frac{h_i}{v}  \,.
\eeqn
When $c_{f,i}\tilde c_{f,i}\neq 0$ or $a_{i}\tilde c_{f,i}\neq 0$, the mass eigenstate $h_i$ couples to both CP-even and CP-odd operators, so the CP symmetry is violated.
The coefficients of $c_{f,i}$, $\tilde c_{f,i}$ and $a_i$ can be derived from the elements of the rotation matrix $\mR$ defined in Eq.~\eqref{eq:rotation_R}, which were also previously obtained in Refs.~\cite{Shu:2013uua, Inoue:2014nva, Chen:2015gaa}.
Here, we summarize their explicit expressions in Table.~\ref{tab:Hcouplings}.
In the special limit of $\beta-\alpha=\pi/2$, the Yukawa couplings and Higgs gauge couplings are determined by the CPV mixing angles of $(\alpha_b\,, \alpha_c)$ and $t_\beta$.
By taking the CPC limit of $\alpha_b=\alpha_c=0$, it is evident that $(h_1\,, h_2)$ have the purely CP-even Yukawa couplings of $c_{f\,,i}$, while $h_3$ has the purely CP-odd Yukawa couplings of $\tilde c_{f\,,i} $.
The previous studies of the collider measurements of the CPV in the Higgs Yukawa couplings can be found in Refs.~\cite{BhupalDev:2007ftb,Harnik:2013aja,Berge:2013jra,Brod:2013cka,Askew:2015mda,Li:2015kxc,Buckley:2015vsa,Berge:2015nua,Han:2016bvf,Hagiwara:2016rdv,Rindani:2016scj,Chen:2017bff,Chen:2017nxp,Azevedo:2017qiz,Hagiwara:2017ban,Goncalves:2018agy,Ma:2018ott,Faroughy:2019ird}.

\newpage

\bibliographystyle{JHEP}
\bibliography{references}

\providecommand{\href}[2]{#2}\begingroup\raggedright\begin{thebibliography}{10}

\bibitem{Aad:2012tfa}
{\scshape ATLAS} collaboration, G.~Aad et~al., \emph{{Observation of a new
  particle in the search for the Standard Model Higgs boson with the ATLAS
  detector at the LHC}},
  \href{http://dx.doi.org/10.1016/j.physletb.2012.08.020}{\emph{Phys. Lett. B}
  {\bf 716} (2012) 1--29}, [\href{http://arxiv.org/abs/1207.7214}{{\tt
  1207.7214}}].

\bibitem{Chatrchyan:2012ufa}
{\scshape CMS} collaboration, S.~Chatrchyan et~al., \emph{{Observation of a New
  Boson at a Mass of 125 GeV with the CMS Experiment at the LHC}},
  \href{http://dx.doi.org/10.1016/j.physletb.2012.08.021}{\emph{Phys. Lett. B}
  {\bf 716} (2012) 30--61}, [\href{http://arxiv.org/abs/1207.7235}{{\tt
  1207.7235}}].

\bibitem{Sakharov:1967dj}
A.~Sakharov, \emph{{Violation of CP Invariance, C asymmetry, and baryon
  asymmetry of the universe}},
  \href{http://dx.doi.org/10.1070/PU1991v034n05ABEH002497}{\emph{Sov. Phys.
  Usp.} {\bf 34} (1991) 392--393}.

\bibitem{Kibble:1976sj}
T.~Kibble, \emph{{Topology of Cosmic Domains and Strings}},
  \href{http://dx.doi.org/10.1088/0305-4470/9/8/029}{\emph{J. Phys. A} {\bf 9}
  (1976) 1387--1398}.

\bibitem{Zurek:1985qw}
W.~Zurek, \emph{{Cosmological Experiments in Superfluid Helium?}},
  \href{http://dx.doi.org/10.1038/317505a0}{\emph{Nature} {\bf 317} (1985)
  505--508}.

\bibitem{Vachaspati:2006zz}
T.~Vachaspati, \emph{{Kinks and domain walls: An introduction to classical and
  quantum solitons}}.
\newblock Cambridge University Press, 4, 2010.

\bibitem{Lee:1974jb}
T.~Lee, \emph{{CP Nonconservation and Spontaneous Symmetry Breaking}},
  \href{http://dx.doi.org/10.1016/0370-1573(74)90020-9}{\emph{Phys. Rept.} {\bf
  9} (1974) 143--177}.

\bibitem{Bao:2009sa}
S.-S. Bao and Y.-L. Wu, \emph{{Neutral Higgs production on LHC in the
  two-Higgs-doublet model with spontaneous CP violation}},
  \href{http://dx.doi.org/10.1103/PhysRevD.81.075020}{\emph{Phys. Rev. D} {\bf
  81} (2010) 075020}, [\href{http://arxiv.org/abs/0907.3606}{{\tt 0907.3606}}].

\bibitem{Ipek:2013iba}
S.~Ipek, \emph{{Perturbative analysis of the electron electric dipole moment
  and CP violation in two-Higgs-doublet models}},
  \href{http://dx.doi.org/10.1103/PhysRevD.89.073012}{\emph{Phys. Rev. D} {\bf
  89} (2014) 073012}, [\href{http://arxiv.org/abs/1310.6790}{{\tt 1310.6790}}].

\bibitem{Grzadkowski:2014ada}
B.~Grzadkowski, O.~M. Ogreid and P.~Osland, \emph{{Measuring CP violation in
  Two-Higgs-Doublet models in light of the LHC Higgs data}},
  \href{http://dx.doi.org/10.1007/JHEP11(2014)084}{\emph{JHEP} {\bf 11} (2014)
  084}, [\href{http://arxiv.org/abs/1409.7265}{{\tt 1409.7265}}].

\bibitem{Eto:2018hhg}
M.~Eto, M.~Kurachi and M.~Nitta, \emph{{Constraints on two Higgs doublet models
  from domain walls}},
  \href{http://dx.doi.org/10.1016/j.physletb.2018.09.002}{\emph{Phys. Lett.}
  {\bf B785} (2018) 447--453}, [\href{http://arxiv.org/abs/1803.04662}{{\tt
  1803.04662}}].

\bibitem{Eto:2018tnk}
M.~Eto, M.~Kurachi and M.~Nitta, \emph{{Non-Abelian strings and domain walls in
  two Higgs doublet models}},
  \href{http://dx.doi.org/10.1007/JHEP08(2018)195}{\emph{JHEP} {\bf 08} (2018)
  195}, [\href{http://arxiv.org/abs/1805.07015}{{\tt 1805.07015}}].

\bibitem{Zeldovich:1974uw}
{\relax Ya}.~B. Zeldovich, I.~{\relax Yu}. Kobzarev and L.~B. Okun,
  \emph{{Cosmological Consequences of the Spontaneous Breakdown of Discrete
  Symmetry}}, {\emph{Zh. Eksp. Teor. Fiz.} {\bf 67} (1974) 3--11}.

\bibitem{Guth:1980zm}
A.~H. Guth, \emph{{The Inflationary Universe: A Possible Solution to the
  Horizon and Flatness Problems}},
  \href{http://dx.doi.org/10.1103/PhysRevD.23.347}{\emph{Adv. Ser. Astrophys.
  Cosmol.} {\bf 3} (1987) 139--148}.

\bibitem{Lazarides:1982tw}
G.~Lazarides and Q.~Shafi, \emph{{Axion Models with No Domain Wall Problem}},
  \href{http://dx.doi.org/10.1016/0370-2693(82)90506-8}{\emph{Phys. Lett. B}
  {\bf 115} (1982) 21--25}.

\bibitem{Barr:1982bb}
S.~M. Barr, D.~Reiss and A.~Zee, \emph{{Families, the Invisible Axion, and
  Domain Walls}},
  \href{http://dx.doi.org/10.1016/0370-2693(82)90331-8}{\emph{Phys. Lett. B}
  {\bf 116} (1982) 227--230}.

\bibitem{Lazarides:2018aev}
G.~Lazarides, M.~Reig, Q.~Shafi, R.~Srivastava and J.~W. Valle,
  \emph{{Spontaneous Breaking of Lepton Number and the Cosmological Domain Wall
  Problem}},
  \href{http://dx.doi.org/10.1103/PhysRevLett.122.151301}{\emph{Phys. Rev.
  Lett.} {\bf 122} (2019) 151301}, [\href{http://arxiv.org/abs/1806.11198}{{\tt
  1806.11198}}].

\bibitem{Vilenkin:1981zs}
A.~Vilenkin, \emph{{Gravitational Field of Vacuum Domain Walls and Strings}},
  \href{http://dx.doi.org/10.1103/PhysRevD.23.852}{\emph{Phys. Rev.} {\bf D23}
  (1981) 852--857}.

\bibitem{Gelmini:1988sf}
G.~B. Gelmini, M.~Gleiser and E.~W. Kolb, \emph{{Cosmology of Biased Discrete
  Symmetry Breaking}},
  \href{http://dx.doi.org/10.1103/PhysRevD.39.1558}{\emph{Phys. Rev. D} {\bf
  39} (1989) 1558}.

\bibitem{Larsson:1996sp}
S.~E. Larsson, S.~Sarkar and P.~L. White, \emph{{Evading the cosmological
  domain wall problem}},
  \href{http://dx.doi.org/10.1103/PhysRevD.55.5129}{\emph{Phys. Rev. D} {\bf
  55} (1997) 5129--5135}, [\href{http://arxiv.org/abs/hep-ph/9608319}{{\tt
  hep-ph/9608319}}].

\bibitem{Hiramatsu:2010yz}
T.~Hiramatsu, M.~Kawasaki and K.~Saikawa, \emph{{Gravitational Waves from
  Collapsing Domain Walls}},
  \href{http://dx.doi.org/10.1088/1475-7516/2010/05/032}{\emph{JCAP} {\bf 1005}
  (2010) 032}, [\href{http://arxiv.org/abs/1002.1555}{{\tt 1002.1555}}].

\bibitem{Kawasaki:2011vv}
M.~Kawasaki and K.~Saikawa, \emph{{Study of gravitational radiation from cosmic
  domain walls}},
  \href{http://dx.doi.org/10.1088/1475-7516/2011/09/008}{\emph{JCAP} {\bf 1109}
  (2011) 008}, [\href{http://arxiv.org/abs/1102.5628}{{\tt 1102.5628}}].

\bibitem{Saikawa:2017hiv}
K.~Saikawa, \emph{{A review of gravitational waves from cosmic domain walls}},
  \href{http://dx.doi.org/10.3390/universe3020040}{\emph{Universe} {\bf 3}
  (2017) 40}, [\href{http://arxiv.org/abs/1703.02576}{{\tt 1703.02576}}].

\bibitem{Zhou:2020ojf}
R.~Zhou, J.~Yang and L.~Bian, \emph{{Gravitational Waves from first-order phase
  transition and domain wall}},
  \href{http://dx.doi.org/10.1007/JHEP04(2020)071}{\emph{JHEP} {\bf 04} (2020)
  071}, [\href{http://arxiv.org/abs/2001.04741}{{\tt 2001.04741}}].

\bibitem{Chen:2020wvu}
N.~Chen, T.~Li and Y.~Wu, \emph{{The gravitational waves from the collapsing
  domain walls in the complex singlet model}},
  \href{http://arxiv.org/abs/2004.10148}{{\tt 2004.10148}}.

\bibitem{Jaeckel:2020mqa}
J.~Jaeckel, S.~Schenk and M.~Spannowsky, \emph{{Probing Dark Matter Clumps,
  Strings and Domain Walls with Gravitational Wave Detectors}},
  \href{http://arxiv.org/abs/2004.13724}{{\tt 2004.13724}}.

\bibitem{AmaroSeoane:2012km}
P.~Amaro-Seoane et~al., \emph{{eLISA/NGO: Astrophysics and cosmology in the
  gravitational-wave millihertz regime}}, {\emph{GW Notes} {\bf 6} (2013)
  4--110}, [\href{http://arxiv.org/abs/1201.3621}{{\tt 1201.3621}}].

\bibitem{AmaroSeoane:2012je}
P.~Amaro-Seoane et~al., \emph{{Low-frequency gravitational-wave science with
  eLISA/NGO}},
  \href{http://dx.doi.org/10.1088/0264-9381/29/12/124016}{\emph{Class. Quant.
  Grav.} {\bf 29} (2012) 124016}, [\href{http://arxiv.org/abs/1202.0839}{{\tt
  1202.0839}}].

\bibitem{Guo:2018npi}
W.-H. Ruan, Z.-K. Guo, R.-G. Cai and Y.-Z. Zhang, \emph{{Taiji Program:
  Gravitational-Wave Sources}},
  \href{http://dx.doi.org/10.1142/S0217751X2050075X}{\emph{Int. J. Mod. Phys.
  A} {\bf 35} (2020) 2050075}, [\href{http://arxiv.org/abs/1807.09495}{{\tt
  1807.09495}}].

\bibitem{Luo:2015ght}
{\scshape TianQin} collaboration, J.~Luo et~al., \emph{{TianQin: a space-borne
  gravitational wave detector}},
  \href{http://dx.doi.org/10.1088/0264-9381/33/3/035010}{\emph{Class. Quant.
  Grav.} {\bf 33} (2016) 035010}, [\href{http://arxiv.org/abs/1512.02076}{{\tt
  1512.02076}}].

\bibitem{Janssen:2014dka}
G.~Janssen et~al., \emph{{Gravitational wave astronomy with the SKA}},
  \href{http://dx.doi.org/10.22323/1.215.0037}{\emph{PoS} {\bf AASKA14} (2015)
  037}, [\href{http://arxiv.org/abs/1501.00127}{{\tt 1501.00127}}].

\bibitem{Kawamura:2011zz}
S.~Kawamura et~al., \emph{{The Japanese space gravitational wave antenna:
  DECIGO}}, \href{http://dx.doi.org/10.1088/0264-9381/28/9/094011}{\emph{Class.
  Quant. Grav.} {\bf 28} (2011) 094011}.

\bibitem{Baron:2013eja}
{\scshape ACME} collaboration, J.~Baron et~al., \emph{{Order of Magnitude
  Smaller Limit on the Electric Dipole Moment of the Electron}},
  \href{http://dx.doi.org/10.1126/science.1248213}{\emph{Science} {\bf 343}
  (2014) 269--272}, [\href{http://arxiv.org/abs/1310.7534}{{\tt 1310.7534}}].

\bibitem{Andreev:2018ayy}
{\scshape ACME} collaboration, V.~Andreev et~al., \emph{{Improved limit on the
  electric dipole moment of the electron}},
  \href{http://dx.doi.org/10.1038/s41586-018-0599-8}{\emph{Nature} {\bf 562}
  (2018) 355--360}.

\bibitem{Graner:2016ses}
B.~Graner, Y.~Chen, E.~Lindahl and B.~Heckel, \emph{{Reduced Limit on the
  Permanent Electric Dipole Moment of Hg199}},
  \href{http://dx.doi.org/10.1103/PhysRevLett.116.161601}{\emph{Phys. Rev.
  Lett.} {\bf 116} (2016) 161601}, [\href{http://arxiv.org/abs/1601.04339}{{\tt
  1601.04339}}].

\bibitem{Bishof:2016uqx}
M.~Bishof et~al., \emph{{Improved limit on the $^{225}$Ra electric dipole
  moment}}, \href{http://dx.doi.org/10.1103/PhysRevC.94.025501}{\emph{Phys.
  Rev. C} {\bf 94} (2016) 025501}, [\href{http://arxiv.org/abs/1606.04931}{{\tt
  1606.04931}}].

\bibitem{Shu:2013uua}
J.~Shu and Y.~Zhang, \emph{{Impact of a CP Violating Higgs Sector: From LHC to
  Baryogenesis}},
  \href{http://dx.doi.org/10.1103/PhysRevLett.111.091801}{\emph{Phys. Rev.
  Lett.} {\bf 111} (2013) 091801}, [\href{http://arxiv.org/abs/1304.0773}{{\tt
  1304.0773}}].

\bibitem{Inoue:2014nva}
S.~Inoue, M.~J. Ramsey-Musolf and Y.~Zhang, \emph{{CP-violating phenomenology
  of flavor conserving two Higgs doublet models}},
  \href{http://dx.doi.org/10.1103/PhysRevD.89.115023}{\emph{Phys. Rev. D} {\bf
  89} (2014) 115023}, [\href{http://arxiv.org/abs/1403.4257}{{\tt 1403.4257}}].

\bibitem{Bian:2014zka}
L.~Bian, T.~Liu and J.~Shu, \emph{{Cancellations Between Two-Loop Contributions
  to the Electron Electric Dipole Moment with a CP-Violating Higgs Sector}},
  \href{http://dx.doi.org/10.1103/PhysRevLett.115.021801}{\emph{Phys. Rev.
  Lett.} {\bf 115} (2015) 021801}, [\href{http://arxiv.org/abs/1411.6695}{{\tt
  1411.6695}}].

\bibitem{Chen:2015gaa}
C.-Y. Chen, S.~Dawson and Y.~Zhang, \emph{{Complementarity of LHC and EDMs for
  Exploring Higgs CP Violation}},
  \href{http://dx.doi.org/10.1007/JHEP06(2015)056}{\emph{JHEP} {\bf 06} (2015)
  056}, [\href{http://arxiv.org/abs/1503.01114}{{\tt 1503.01114}}].

\bibitem{Bian:2016zba}
L.~Bian and N.~Chen, \emph{{Cancellation mechanism in the predictions of
  electric dipole moments}},
  \href{http://dx.doi.org/10.1103/PhysRevD.95.115029}{\emph{Phys. Rev. D} {\bf
  95} (2017) 115029}, [\href{http://arxiv.org/abs/1608.07975}{{\tt
  1608.07975}}].

\bibitem{Egana-Ugrinovic:2018fpy}
D.~Egana-Ugrinovic and S.~Thomas, \emph{{Higgs Boson Contributions to the
  Electron Electric Dipole Moment}},
  \href{http://arxiv.org/abs/1810.08631}{{\tt 1810.08631}}.

\bibitem{Chun:2019oix}
E.~J. Chun, J.~Kim and T.~Mondal, \emph{{Electron EDM and Muon anomalous
  magnetic moment in Two-Higgs-Doublet Models}},
  \href{http://dx.doi.org/10.1007/JHEP12(2019)068}{\emph{JHEP} {\bf 12} (2019)
  068}, [\href{http://arxiv.org/abs/1906.00612}{{\tt 1906.00612}}].

\bibitem{Cheung:2020ugr}
K.~Cheung, A.~Jueid, Y.-N. Mao and S.~Moretti, \emph{{The 2-Higgs-Doublet Model
  with Soft CP-violation Confronting Electric Dipole Moments and Colliders}},
  \href{http://arxiv.org/abs/2003.04178}{{\tt 2003.04178}}.

\bibitem{Kanemura:2020ibp}
S.~Kanemura, M.~Kubota and K.~Yagyu, \emph{{Aligned CP-violating Higgs sector
  canceling the electric dipole moment}},
  \href{http://arxiv.org/abs/2004.03943}{{\tt 2004.03943}}.

\bibitem{Cesarotti:2018huy}
C.~Cesarotti, Q.~Lu, Y.~Nakai, A.~Parikh and M.~Reece, \emph{{Interpreting the
  Electron EDM Constraint}},
  \href{http://dx.doi.org/10.1007/JHEP05(2019)059}{\emph{JHEP} {\bf 05} (2019)
  059}, [\href{http://arxiv.org/abs/1810.07736}{{\tt 1810.07736}}].

\bibitem{Davidson:2005cw}
S.~Davidson and H.~E. Haber, \emph{{Basis-independent methods for the
  two-Higgs-doublet model}},
  \href{http://dx.doi.org/10.1103/PhysRevD.72.099902,
  10.1103/PhysRevD.72.035004}{\emph{Phys. Rev.} {\bf D72} (2005) 035004},
  [\href{http://arxiv.org/abs/hep-ph/0504050}{{\tt hep-ph/0504050}}].

\bibitem{Gunion:2005ja}
J.~F. Gunion and H.~E. Haber, \emph{{Conditions for CP-violation in the general
  two-Higgs-doublet model}},
  \href{http://dx.doi.org/10.1103/PhysRevD.72.095002}{\emph{Phys. Rev.} {\bf
  D72} (2005) 095002}, [\href{http://arxiv.org/abs/hep-ph/0506227}{{\tt
  hep-ph/0506227}}].

\bibitem{Ivanov:2006yq}
I.~P. Ivanov, \emph{{Minkowski space structure of the Higgs potential in
  2HDM}}, \href{http://dx.doi.org/10.1103/PhysRevD.76.039902,
  10.1103/PhysRevD.75.035001}{\emph{Phys. Rev.} {\bf D75} (2007) 035001},
  [\href{http://arxiv.org/abs/hep-ph/0609018}{{\tt hep-ph/0609018}}].

\bibitem{Ivanov:2007de}
I.~P. Ivanov, \emph{{Minkowski space structure of the Higgs potential in 2HDM.
  II. Minima, symmetries, and topology}},
  \href{http://dx.doi.org/10.1103/PhysRevD.77.015017}{\emph{Phys. Rev. D} {\bf
  77} (2008) 015017}, [\href{http://arxiv.org/abs/0710.3490}{{\tt 0710.3490}}].

\bibitem{Branco:2011iw}
G.~Branco, P.~Ferreira, L.~Lavoura, M.~Rebelo, M.~Sher and J.~P. Silva,
  \emph{{Theory and phenomenology of two-Higgs-doublet models}},
  \href{http://dx.doi.org/10.1016/j.physrep.2012.02.002}{\emph{Phys. Rept.}
  {\bf 516} (2012) 1--102}, [\href{http://arxiv.org/abs/1106.0034}{{\tt
  1106.0034}}].

\bibitem{Battye:2011jj}
R.~A. Battye, G.~D. Brawn and A.~Pilaftsis, \emph{{Vacuum Topology of the Two
  Higgs Doublet Model}},
  \href{http://dx.doi.org/10.1007/JHEP08(2011)020}{\emph{JHEP} {\bf 08} (2011)
  020}, [\href{http://arxiv.org/abs/1106.3482}{{\tt 1106.3482}}].

\bibitem{Grzadkowski:2013rza}
B.~Grzadkowski, O.~M. Ogreid and P.~Osland, \emph{{Diagnosing CP properties of
  the 2HDM}}, \href{http://dx.doi.org/10.1007/JHEP01(2014)105}{\emph{JHEP} {\bf
  01} (2014) 105}, [\href{http://arxiv.org/abs/1309.6229}{{\tt 1309.6229}}].

\bibitem{Grzadkowski:2009bt}
B.~Grzadkowski, O.~Ogreid and P.~Osland, \emph{{Natural Multi-Higgs Model with
  Dark Matter and CP Violation}},
  \href{http://dx.doi.org/10.1103/PhysRevD.80.055013}{\emph{Phys. Rev. D} {\bf
  80} (2009) 055013}, [\href{http://arxiv.org/abs/0904.2173}{{\tt 0904.2173}}].

\bibitem{Arhrib:2000is}
A.~Arhrib, \emph{{Unitarity constraints on scalar parameters of the standard
  and two Higgs doublets model}},  in \emph{{Workshop on Noncommutative
  Geometry, Superstrings and Particle Physics}}, 12, 2000.
\newblock \href{http://arxiv.org/abs/hep-ph/0012353}{{\tt hep-ph/0012353}}.

\bibitem{Kanemura:2015ska}
S.~Kanemura and K.~Yagyu, \emph{{Unitarity bound in the most general two Higgs
  doublet model}},
  \href{http://dx.doi.org/10.1016/j.physletb.2015.10.047}{\emph{Phys. Lett. B}
  {\bf 751} (2015) 289--296}, [\href{http://arxiv.org/abs/1509.06060}{{\tt
  1509.06060}}].

\bibitem{brawn_SymmetriesTopologicalDefects_2011}
G.~Brawn, \emph{Symmetries and {Topological} {Defects} of the {Two} {Higgs}
  {Doublet} {Model}}.
\newblock Ph.{D}., University of Manchester, UK, 2011.

\bibitem{Wainwright:2011kj}
C.~L. Wainwright, \emph{{CosmoTransitions: Computing Cosmological Phase
  Transition Temperatures and Bubble Profiles with Multiple Fields}},
  \href{http://dx.doi.org/10.1016/j.cpc.2012.04.004}{\emph{Comput. Phys.
  Commun.} {\bf 183} (2012) 2006--2013},
  [\href{http://arxiv.org/abs/1109.4189}{{\tt 1109.4189}}].

\bibitem{Stauffer:1978kr}
D.~Stauffer, \emph{{Scaling theory of percolation clusters}},
  \href{http://dx.doi.org/10.1016/0370-1573(79)90060-7}{\emph{Phys. Rept.} {\bf
  54} (1979) 1--74}.

\bibitem{Kawasaki:2004yh}
M.~Kawasaki, K.~Kohri and T.~Moroi, \emph{{Hadronic decay of late - decaying
  particles and Big-Bang Nucleosynthesis}},
  \href{http://dx.doi.org/10.1016/j.physletb.2005.08.045}{\emph{Phys. Lett. B}
  {\bf 625} (2005) 7--12}, [\href{http://arxiv.org/abs/astro-ph/0402490}{{\tt
  astro-ph/0402490}}].

\bibitem{Kawasaki:2004qu}
M.~Kawasaki, K.~Kohri and T.~Moroi, \emph{{Big-Bang nucleosynthesis and
  hadronic decay of long-lived massive particles}},
  \href{http://dx.doi.org/10.1103/PhysRevD.71.083502}{\emph{Phys. Rev. D} {\bf
  71} (2005) 083502}, [\href{http://arxiv.org/abs/astro-ph/0408426}{{\tt
  astro-ph/0408426}}].

\bibitem{Hiramatsu:2013qaa}
T.~Hiramatsu, M.~Kawasaki and K.~Saikawa, \emph{{On the estimation of
  gravitational wave spectrum from cosmic domain walls}},
  \href{http://dx.doi.org/10.1088/1475-7516/2014/02/031}{\emph{JCAP} {\bf 1402}
  (2014) 031}, [\href{http://arxiv.org/abs/1309.5001}{{\tt 1309.5001}}].

\bibitem{Gleiser:1998na}
M.~Gleiser and R.~Roberts, \emph{{Gravitational waves from collapsing vacuum
  domains}}, \href{http://dx.doi.org/10.1103/PhysRevLett.81.5497}{\emph{Phys.
  Rev. Lett.} {\bf 81} (1998) 5497--5500},
  [\href{http://arxiv.org/abs/astro-ph/9807260}{{\tt astro-ph/9807260}}].

\bibitem{Hoogeveen:1990cb}
F.~Hoogeveen, \emph{{The Standard Model Prediction for the Electric Dipole
  Moment of the Electron}},
  \href{http://dx.doi.org/10.1016/0550-3213(90)90182-D}{\emph{Nucl. Phys. B}
  {\bf 341} (1990) 322--340}.

\bibitem{Pospelov:1991zt}
M.~Pospelov and I.~Khriplovich, \emph{{Electric dipole moment of the W boson
  and the electron in the Kobayashi-Maskawa model}}, {\emph{Sov. J. Nucl.
  Phys.} {\bf 53} (1991) 638--640}.

\bibitem{Pospelov:2013sca}
M.~Pospelov and A.~Ritz, \emph{{CKM benchmarks for electron electric dipole
  moment experiments}},
  \href{http://dx.doi.org/10.1103/PhysRevD.89.056006}{\emph{Phys. Rev. D} {\bf
  89} (2014) 056006}, [\href{http://arxiv.org/abs/1311.5537}{{\tt 1311.5537}}].

\bibitem{Yamaguchi:2020dsy}
Y.~Yamaguchi and N.~Yamanaka, \emph{{Quark level and hadronic contributions to
  the electric dipole moment of charged leptons in the standard model}},
  \href{http://arxiv.org/abs/2006.00281}{{\tt 2006.00281}}.

\bibitem{Barr:1990vd}
S.~M. Barr and A.~Zee, \emph{{Electric Dipole Moment of the Electron and of the
  Neutron}}, \href{http://dx.doi.org/10.1103/PhysRevLett.65.21}{\emph{Phys.
  Rev. Lett.} {\bf 65} (1990) 21--24}.

\bibitem{La:1993je}
H.~La, \emph{{Vortex solutions in two Higgs systems and tan Beta}},
  \href{http://arxiv.org/abs/hep-ph/9302220}{{\tt hep-ph/9302220}}.

\bibitem{Earnshaw:1993yu}
M.~A. Earnshaw and M.~James, \emph{{Stability of two doublet electroweak
  strings}}, \href{http://dx.doi.org/10.1103/PhysRevD.48.5818}{\emph{Phys.
  Rev.} {\bf D48} (1993) 5818--5826},
  [\href{http://arxiv.org/abs/hep-ph/9308223}{{\tt hep-ph/9308223}}].

\bibitem{Bimonte:1994qh}
G.~Bimonte and G.~Lozano, \emph{{Vortex solutions in two Higgs doublet
  systems}}, \href{http://dx.doi.org/10.1016/0370-2693(94)91321-8}{\emph{Phys.
  Lett.} {\bf B326} (1994) 270--275},
  [\href{http://arxiv.org/abs/hep-ph/9401313}{{\tt hep-ph/9401313}}].

\bibitem{Eto:2019hhf}
M.~Eto, Y.~Hamada, M.~Kurachi and M.~Nitta, \emph{{Topological Nambu monopole
  in two Higgs doublet models}},
  \href{http://dx.doi.org/10.1016/j.physletb.2020.135220}{\emph{Phys. Lett. B}
  {\bf 802} (2020) 135220}, [\href{http://arxiv.org/abs/1904.09269}{{\tt
  1904.09269}}].

\bibitem{Eto:2020hjb}
M.~Eto, Y.~Hamada, M.~Kurachi and M.~Nitta, \emph{{Dynamics of Nambu monopole
  in two Higgs doublet models -- Cosmological Monopole Collider --}},
  \href{http://arxiv.org/abs/2003.08772}{{\tt 2003.08772}}.

\bibitem{Rubakov:1981rg}
V.~Rubakov, \emph{{Superheavy Magnetic Monopoles and Proton Decay}},
  {\emph{JETP Lett.} {\bf 33} (1981) 644--646}.

\bibitem{Callan:1982au}
J.~Callan, Curtis~G., \emph{{Dyon-Fermion Dynamics}},
  \href{http://dx.doi.org/10.1103/PhysRevD.26.2058}{\emph{Phys. Rev. D} {\bf
  26} (1982) 2058--2068}.

\bibitem{Callan:1982ac}
J.~Callan, Curtis~G., \emph{{Monopole Catalysis of Baryon Decay}},
  \href{http://dx.doi.org/10.1016/0550-3213(83)90677-6}{\emph{Nucl. Phys. B}
  {\bf 212} (1983) 391--400}.

\bibitem{Cline:2020jre}
J.~M. Cline and K.~Kainulainen, \emph{{Electroweak baryogenesis at high bubble
  wall velocities}},
  \href{http://dx.doi.org/10.1103/PhysRevD.101.063525}{\emph{Phys. Rev. D} {\bf
  101} (2020) 063525}, [\href{http://arxiv.org/abs/2001.00568}{{\tt
  2001.00568}}].

\bibitem{Zhou:2020xqi}
R.~Zhou and L.~Bian, \emph{{Baryon asymmetry and detectable Gravitational Waves
  from Electroweak phase transition}},
  \href{http://arxiv.org/abs/2001.01237}{{\tt 2001.01237}}.

\bibitem{Xie:2020bkl}
K.-P. Xie, Y.~Wu and L.~Bian, \emph{{Electroweak baryogenesis and gravitational
  waves in a composite Higgs model with high dimensional fermion
  representations}},  \href{http://arxiv.org/abs/2005.13552}{{\tt 2005.13552}}.

\bibitem{WahabElKaffas:2007xd}
A.~Wahab El~Kaffas, P.~Osland and O.~M. Ogreid, \emph{{Constraining the
  Two-Higgs-Doublet-Model parameter space}},
  \href{http://dx.doi.org/10.1103/PhysRevD.76.095001}{\emph{Phys. Rev. D} {\bf
  76} (2007) 095001}, [\href{http://arxiv.org/abs/0706.2997}{{\tt 0706.2997}}].

\bibitem{Khater:2003wq}
W.~Khater and P.~Osland, \emph{{CP violation in top quark production at the LHC
  and two Higgs doublet models}},
  \href{http://dx.doi.org/10.1016/S0550-3213(03)00300-6}{\emph{Nucl. Phys.}
  {\bf B661} (2003) 209--234}, [\href{http://arxiv.org/abs/hep-ph/0302004}{{\tt
  hep-ph/0302004}}].

\bibitem{BhupalDev:2007ftb}
P.~Bhupal~Dev, A.~Djouadi, R.~Godbole, M.~Muhlleitner and S.~Rindani,
  \emph{{Determining the CP properties of the Higgs boson}},
  \href{http://dx.doi.org/10.1103/PhysRevLett.100.051801}{\emph{Phys. Rev.
  Lett.} {\bf 100} (2008) 051801}, [\href{http://arxiv.org/abs/0707.2878}{{\tt
  0707.2878}}].

\bibitem{Harnik:2013aja}
R.~Harnik, A.~Martin, T.~Okui, R.~Primulando and F.~Yu, \emph{{Measuring CP
  Violation in $h \to \tau^+ \tau^-$ at Colliders}},
  \href{http://dx.doi.org/10.1103/PhysRevD.88.076009}{\emph{Phys. Rev. D} {\bf
  88} (2013) 076009}, [\href{http://arxiv.org/abs/1308.1094}{{\tt 1308.1094}}].

\bibitem{Berge:2013jra}
S.~Berge, W.~Bernreuther and H.~Spiesberger, \emph{{Higgs CP properties using
  the $\tau$ decay modes at the ILC}},
  \href{http://dx.doi.org/10.1016/j.physletb.2013.11.006}{\emph{Phys. Lett. B}
  {\bf 727} (2013) 488--495}, [\href{http://arxiv.org/abs/1308.2674}{{\tt
  1308.2674}}].

\bibitem{Brod:2013cka}
J.~Brod, U.~Haisch and J.~Zupan, \emph{{Constraints on CP-violating Higgs
  couplings to the third generation}},
  \href{http://dx.doi.org/10.1007/JHEP11(2013)180}{\emph{JHEP} {\bf 11} (2013)
  180}, [\href{http://arxiv.org/abs/1310.1385}{{\tt 1310.1385}}].

\bibitem{Askew:2015mda}
A.~Askew, P.~Jaiswal, T.~Okui, H.~B. Prosper and N.~Sato, \emph{{Prospect for
  measuring the CP phase in the $h\tau\tau$ coupling at the LHC}},
  \href{http://dx.doi.org/10.1103/PhysRevD.91.075014}{\emph{Phys. Rev. D} {\bf
  91} (2015) 075014}, [\href{http://arxiv.org/abs/1501.03156}{{\tt
  1501.03156}}].

\bibitem{Li:2015kxc}
G.~Li, H.-R. Wang and S.-h. Zhu, \emph{{Probing CP-violating $h\bar{t}t$
  coupling in $e^{+}e^{-}\rightarrow h \gamma$}},
  \href{http://dx.doi.org/10.1103/PhysRevD.93.055038}{\emph{Phys. Rev. D} {\bf
  93} (2016) 055038}, [\href{http://arxiv.org/abs/1506.06453}{{\tt
  1506.06453}}].

\bibitem{Buckley:2015vsa}
M.~R. Buckley and D.~Goncalves, \emph{{Boosting the Direct CP Measurement of
  the Higgs-Top Coupling}},
  \href{http://dx.doi.org/10.1103/PhysRevLett.116.091801}{\emph{Phys. Rev.
  Lett.} {\bf 116} (2016) 091801}, [\href{http://arxiv.org/abs/1507.07926}{{\tt
  1507.07926}}].

\bibitem{Berge:2015nua}
S.~Berge, W.~Bernreuther and S.~Kirchner, \emph{{Prospects of constraining the
  Higgs boson's CP nature in the tau decay channel at the LHC}},
  \href{http://dx.doi.org/10.1103/PhysRevD.92.096012}{\emph{Phys. Rev. D} {\bf
  92} (2015) 096012}, [\href{http://arxiv.org/abs/1510.03850}{{\tt
  1510.03850}}].

\bibitem{Han:2016bvf}
T.~Han, S.~Mukhopadhyay, B.~Mukhopadhyaya and Y.~Wu, \emph{{Measuring the CP
  property of Higgs coupling to tau leptons in the VBF channel at the LHC}},
  \href{http://dx.doi.org/10.1007/JHEP05(2017)128}{\emph{JHEP} {\bf 05} (2017)
  128}, [\href{http://arxiv.org/abs/1612.00413}{{\tt 1612.00413}}].

\bibitem{Hagiwara:2016rdv}
K.~Hagiwara, K.~Ma and H.~Yokoya, \emph{{Probing CP violation in $e^{+}e^{-}$
  production of the Higgs boson and toponia}},
  \href{http://dx.doi.org/10.1007/JHEP06(2016)048}{\emph{JHEP} {\bf 06} (2016)
  048}, [\href{http://arxiv.org/abs/1602.00684}{{\tt 1602.00684}}].

\bibitem{Rindani:2016scj}
S.~D. Rindani, P.~Sharma and A.~Shivaji, \emph{{Unraveling the CP phase of
  top-Higgs coupling in associated production at the LHC}},
  \href{http://dx.doi.org/10.1016/j.physletb.2016.08.002}{\emph{Phys. Lett. B}
  {\bf 761} (2016) 25--30}, [\href{http://arxiv.org/abs/1605.03806}{{\tt
  1605.03806}}].

\bibitem{Chen:2017bff}
X.~Chen and Y.~Wu, \emph{{Search for CP violation effects in the $h\to
  \tau\tau$ decay with $e^+e^-$ colliders}},
  \href{http://dx.doi.org/10.1140/epjc/s10052-017-5258-y}{\emph{Eur. Phys. J.
  C} {\bf 77} (2017) 697}, [\href{http://arxiv.org/abs/1703.04855}{{\tt
  1703.04855}}].

\bibitem{Chen:2017nxp}
X.~Chen and Y.~Wu, \emph{{Probing the CP-Violation effects in the $h\tau\tau$
  coupling at the LHC}},
  \href{http://dx.doi.org/10.1016/j.physletb.2019.01.038}{\emph{Phys. Lett. B}
  {\bf 790} (2019) 332--338}, [\href{http://arxiv.org/abs/1708.02882}{{\tt
  1708.02882}}].

\bibitem{Azevedo:2017qiz}
D.~Azevedo, A.~Onofre, F.~Filthaut and R.~Gonçalo, \emph{{CP tests of Higgs
  couplings in $t\bar{t}h$ semileptonic events at the LHC}},
  \href{http://dx.doi.org/10.1103/PhysRevD.98.033004}{\emph{Phys. Rev. D} {\bf
  98} (2018) 033004}, [\href{http://arxiv.org/abs/1711.05292}{{\tt
  1711.05292}}].

\bibitem{Hagiwara:2017ban}
K.~Hagiwara, H.~Yokoya and Y.-J. Zheng, \emph{{Probing the CP properties of top
  Yukawa coupling at an $e^+e^-$ collider}},
  \href{http://dx.doi.org/10.1007/JHEP02(2018)180}{\emph{JHEP} {\bf 02} (2018)
  180}, [\href{http://arxiv.org/abs/1712.09953}{{\tt 1712.09953}}].

\bibitem{Goncalves:2018agy}
D.~Gonçalves, K.~Kong and J.~H. Kim, \emph{{Probing the top-Higgs Yukawa CP
  structure in dileptonic $ t\overline{t}h $ with M$_{2}$-assisted
  reconstruction}},
  \href{http://dx.doi.org/10.1007/JHEP06(2018)079}{\emph{JHEP} {\bf 06} (2018)
  079}, [\href{http://arxiv.org/abs/1804.05874}{{\tt 1804.05874}}].

\bibitem{Ma:2018ott}
K.~Ma, \emph{{Enhancing $CP$ Measurement of the Yukawa Interactions of
  Top-Quark at $e^{-}e^{+}$ Collider}},
  \href{http://dx.doi.org/10.1016/j.physletb.2019.134928}{\emph{Phys. Lett. B}
  {\bf 797} (2019) 134928}, [\href{http://arxiv.org/abs/1809.07127}{{\tt
  1809.07127}}].

\bibitem{Faroughy:2019ird}
D.~A. Faroughy, J.~F. Kamenik, N.~Ko\v~snik and A.~Smolkovi\v~c, \emph{{Probing
  the $CP$ nature of the top quark Yukawa at hadron colliders}},
  \href{http://dx.doi.org/10.1007/JHEP02(2020)085}{\emph{JHEP} {\bf 02} (2020)
  085}, [\href{http://arxiv.org/abs/1909.00007}{{\tt 1909.00007}}].

\end{thebibliography}\endgroup
\end{document}